\documentstyle[twocolumn,mathptm]{mn}

\renewcommand{\d}{{\rm d}}

\input{psfig.sty}

\begin{document}

\title{Cluster cross sections for strong lensing: analytic and
numerical lens models}

\author[Meneghetti et al.]
  {Massimo Meneghetti$^{1,2}$, Matthias Bartelmann$^2$, Lauro
   Moscardini$^{1,3}$\\
   $^1$Dipartimento di Astronomia, Universit\`a di Padova, vicolo
   dell'Osservatorio 2, I--35122 Padova, Italy\\
   $^2$Max-Planck-Institut f\"ur Astrophysik, P.O.~Box 1317, D--85748
   Garching, Germany\\
   $^3$Present address: Dipartimento di Astronomia,
   Universit\`a di Bologna, via Ranzani 1, I--40127 Bologna, Italy}

\date{Accepted 2002 ???? ???; Received 2002 ???? ???;
in original form 2002 ???? ??}

\maketitle

\begin{abstract}

The statistics of gravitationally lensed arcs was recognised earlier
as a potentially powerful cosmological probe. However, while fully
numerical models find orders of magnitude difference between the arc
probabilities in different cosmological models, analytic models tend
to find markedly different results. We introduce in this paper an
analytic cluster lens model which improves upon existing analytic
models in four ways. (1) We use the more realistic Navarro-Frenk-White
profile instead of singular isothermal spheres, (2) we include the
effect of cosmology on the compactness of the lenses, (3) we use
elliptical instead of axially symmetric lenses, and (4) we take the
intrinsic ellipticity of sources into account. While these
improvements to the analytic model lead to a pronounced increase of
the arc probability, comparisons with numerical models of the same
virial mass demonstrate that the analytic models still fall short by a
substantial margin of reproducing the results obtained with numerical
models. Using multipole expansions of cluster mass distributions, we
show that the remaining discrepancy can be attributed to substructure
inside clusters and tidal fields contributed by the cluster
surroundings, effects that cannot reasonably and reliably be mimicked
in analytic models.

\end{abstract}

\section{Introduction}

Many authors pointed out that the statistics of gravitationally lensed
arcs in galaxy clusters may be a powerful tool for constraining
cosmological models. They are rare events caused by a highly nonlinear
effect in cluster cores, and are thus not only sensitive to the
cosmologically highly variable number density of galaxy clusters, but
also to their internal structure.

The expected number of {\em giant\/} arcs, usually defined as arcs
with a length-to-width ratio exceeding ten and apparent $B$-magnitude
less than $22.5$ \cite{wu93}, changes by orders of magnitude between
low- and high-density universes according to the numerical models
described in Bartelmann et al. (1998).  They use the ray-tracing
technique for studying gravitational lensing by galaxy cluster models
taken from N-body simulations (see also Bartelmann \& Weiss 1994;
Bartelmann 1995; Bartelmann, Steinmetz \& Weiss 1995; Meneghetti et
al. 2000, 2001). This allows the most realistic description of the
cluster lenses because all effects which could play an important role
for the lensing phenomena are by construction taken into
account. Cluster asymmetries, substructure, and the tidal field of the
surrounding matter distribution are known to have substantial effects
on arc statistics, and they are automatically included if the cluster
simulations are suitably designed.

In order to extract useful cosmological constraints from arc
statistics, it is essential to perform simulations on a fine grid in
the cosmological parameter space. However, given the long computation
times required for full numerical simulations of cluster lensing, it
is currently not feasible to perform such simulations for sufficiently
many combinations of the essential cosmological parameters, i.e.~the
matter density parameter $\Omega_0$ and the cosmological constant
$\Omega_\Lambda$. Therefore, conclusions can so far be drawn only for
discrete points in the $\Omega_0$--$\Omega_\Lambda$ plane.

In a conceptually different approach, simple analytic, axially
symmetric models have been used for describing the density profiles of
cluster lenses (see e.g. Wu \& Mao 1996; Cooray, Quashnock \& Miller
1999; Molikawa et al.  1999; Kaufmann \& Straumann 2000; Oguri, Taraya
\& Suto 2001; Oguri et al. 2002; Molikawa \& Hattori 2001).  This
method of investigation has the advantage that the computation of the
probability for arcs satisfying a specified property is fast and can
easily be performed for a continuous and wide range of cosmological
parameters, because the lensing properties of these models are
perfectly known and fully described by analytic formulae. However,
important effects like substructures or asymmetries in the matter
distribution can at best be taken into account at an approximate
level, and the correspondence between analytic and numerical models
remains unclear. While the analytic studies by Cooray et al. (1999)
and Kaufmann \& Straumann (2000) find similar results as the numerical
simulations regarding the sensitivity of arc statistics to the cosmic
density parameter, their results are almost insensitive to the
cosmological constant, in marked contrast to Bartelmann et al. (1998),
who found order-of-magnitude changes in the arc cross sections between
low-density models with and without a cosmological constant.

In this paper, we investigate whether the results of the analytic and
numerical approaches can be reconciled using a more realistic analytic
lens model. Previous analytic studies of arc statistics commonly used
the singular isothermal sphere (hereafter SIS), which is
computationally convenient, but has an unrealistic density profile and
does not naturally reflect the theoretically expected and numerically
demonstrated variations in halo concentration. Instead, we model
cluster lenses as haloes with the density profile found by Navarro,
Frenk \& White (1997; hereafter NFW) in high-resolution simulations of
haloes with a wide range of masses. The profile is flatter than
isothermal near the core, and steeper outside, which leads to
qualitatively and quantitatively different lensing properties. We take
the change of halo concentration with halo mass and cosmology into
account, we distort the lensing potential elliptically in order to
mimic cluster asymmetries, and we adapt the ellipticity by fitting
numerical cluster models.  We then compare the efficiency for
producing arcs with a minimal length-to-width ratio of the elliptical
NFW lenses with fully numerically simulated cluster lenses, and with
the singular isothermal spheres for reference.

The plan of the paper is as follows.  In Sect.~\ref{section:cross}, we
define the lensing cross section of both numerical and analytic
models. In particular, in Sect.~\ref{section:numcross} we describe the
ray-tracing simulations, and we show the lensing cross sections
produced by a sample of five galaxy clusters simulated in three
different cosmological models. In Sect.~\ref{section:sphcross}, we
deal with the lensing cross sections of axially symmetric analytic
models and discuss the differences between SIS and NFW models. In
Sect.~\ref{section:ellsrc}, we show how the lensing cross sections
change by assuming elliptical instead of circular sources. In
Sect.~\ref{section:ellcross}, we discuss lensing by elliptically
distorted NFW models and compare their arc cross sections with those
of axially symmetric NFW lenses. In Sect.~\ref{section:compa} we
compare the analytic and the numerical models. Our conclusions are
presented in Sect.~\ref{section:results}.

\section{Cross sections for long and thin arcs}
\label{section:cross}

The efficiency of a given lens for producing arcs with specified
properties can be quantified by means of suitably defined lensing
cross sections. They are defined as the areas on the source plane
where sources must be located in order to be imaged as arcs satisfying
the required conditions on, e.g., size, shape, magnitude and the
like. We focus in this paper on the cross sections of clusters for
producing long and thin arcs, i.e.~for arcs whose length-to-width
ratio $L/W$ exceeds a given minimum value $(L/W)_{\rm min}$. In the
following subsections, we describe our method for computing strong
lensing cross sections of numerically and analytically modelled galaxy
clusters.

\subsection{Numerical models}
\label{section:numcross}

For comparison to the analytic models, we compute the strong lensing
cross sections of five numerically simulated cluster-sized dark-matter
haloes, kindly made available by the GIF collaboration
\cite{kauffmann99}. The same clusters have been used by Bartelmann et
al.  (1998). They were obtained from $N$-body simulations performed in
the framework of three different cosmological models. These are an
Einstein-de Sitter model (hereafter SCDM); a flat, low-density
universe with a matter density parameter $\Omega_0=0.3$ and a
cosmological constant $\Omega_\Lambda=0.7$ ($\Lambda$CDM); and an
open, low-density model with $\Omega_0=0.3$ and $\Omega_\Lambda=0$
(OCDM). The initial matter density in these models is perturbed about
the mean according to a CDM power spectrum \cite{bond84} with
primordial spectral index $n=1$, normalised such that the local
abundance of massive galaxy clusters is reproduced (e.g.~Viana \&
Liddle 1996). The complete list of cosmological parameters in these
simulations is given in Bartelmann et al. (1998). The virial masses of
the clusters at redshift $z=0$ range between $\sim 5 \times 10^{14}
M_{\odot}/h$ and $\sim 2 \times 10^{15} M_{\odot}/h$. The masses of
individual particles are $1.0 \times 10^{10} h^{-1} M_{\odot}$ and
$1.4 \times 10^{10} h^{-1} M_{\odot}$ for the high- and the
low-$\Omega_0$ models, respectively.

The lensing properties of these clusters are studied using a
ray-shooting technique which was described in detail elsewhere
(Bartelmann et al. 1998; Meneghetti et al. 2000, 2001). Here, we only
briefly discuss some parameters used in these computations, and refer
the reader to the cited papers for a more complete description.

Since strong lensing only occurs near the very cluster centres, we
select areas of $1.5\,h^{-1}\,{\rm Mpc}$ comoving side length centred
on the lens centres. For all reasonable combinations of lens and
source redshifts, this region is large enough to encompass the
critical curves of the cluster lens, close to which large arcs form. A
bundle of $2048\times2048$ rays is shot through a regular grid on the
lens plane, covering the studied region, and their paths to the source
plane are traced. We consider different snapshots of the cluster
simulation at selected redshifts between $z_{\rm l}=0$ and $z_{\rm
l}=1$. For each snapshot, three projections of each simulated cluster
on orthogonal planes are used for performing the lensing
simulations. The typical angular resolution at $z\approx 0.3$ is 0.18
arcsec, assuming the $\Lambda$CDM model.

The sources are assumed to all lie on one plane at redshift $z_{\rm
s}=1$. Although real galaxy sources are obviously distributed in
redshift, putting them all at the same redshift is acceptable because
first, observations show that most of the sources which experience
strong lensing effects are generally at redshifts near unity and
second, the critical surface density changes very little with source
redshift, unless it is very close to the lens redshift. Moreover, we
assume sources to be elliptical, with their axis ratios $b/a$
uniformly distributed in the range $0.5$--$1.0$. An initial set of
sources is placed on a regular grid covering the source plane, and
additional sources are added on sub-grids whose resolution is
iteratively increased towards the lens caustics. Thus, sources are
placed on an adaptive hierarchy of grids in order to improve the
numerical efficiency of the method and to increase the probability for
finding long and thin arcs. For correcting the statistics of the
numerically simulated arcs, we assign to each source and all of its
images a statistical weight $w$ which is inversely proportional to the
squared resolution of the sub-grid on which the corresponding source
was placed. The finer the grid resolution, the lower is the
statistical weight of its images; see also Eq.~(\ref{equation:ncross})
below.

Using the ray-tracing technique, we reconstruct the images of the
background sources and measure their length, width and curvature
radius. Our technique for image detection and classification was
described in detail by Bartelmann \& Weiss (1994) and adopted by
Bartelmann et al. (1998) and Meneghetti et al. (2000, 2001). It
results in a catalogue of simulated images which is subsequently
analysed statistically.

Each source is taken to represent a fraction of the source plane. The
cells of the sub-grid with the highest resolution have area $A$, and
the sources placed on its grid points are given a statistical weight
of unity. The absolute lensing cross sections for a specified image
property are then determined by counting the statistical weights of
the sources whose images have the required property. Specifically, we
search for sources with a length-to-width ratio exceeding a threshold
$(L/W)_{\rm min}$. If a source has multiple images with
$(L/W)\ge(L/W)_{\rm min}$, we multiply its statistical weight by the
number of such images. Therefore, the lensing cross section is
\begin{equation}
\label{equation:ncross}
  \sigma_{(L/W)_{\rm min}}=A\,\sum_i\,W_i w_i n_i \ ,
\end{equation}
where $W_i$ is unity if the $i$-th source has images with
$(L/W)\ge(L/W)_{\rm min}$ and zero otherwise, $n_i$ is the number of
images of the $i$-th source satisfying the required condition, and
$w_i$ is the statistical weight of the source.

\subsection{Axially symmetric models}
\label{section:sphcross}

For sufficiently simple, axially symmetric lens models, the strong
lensing cross sections can be also computed in an analytic or
semi-analytic way. We consider two such models, the singular
isothermal sphere (SIS) for reference, and haloes with the NFW density
profile \cite{navarro97}. The SIS has been very widely used in many
previous studies on arc statistics because of its computational
simplicity. The NFW density profile fits the results of highly
resolved numerical halo simulations and thus provides a much more
realistic description of cluster density profiles than the SIS
profile. We focus on the NFW profile here, but give some results for
SIS lenses for later comparison.

The SIS density profile is given by
\begin{equation}
  \rho(r)=\frac{\sigma_v^2}{(2 \pi G r^2)} \ ,
\end{equation}
where $\sigma_v$ is the velocity dispersion. The NFW density profile
is
\begin{equation}
  \rho(r)=\frac{\rho_{\rm s}}{(r/r_{\rm s})(1+r/r_{\rm s})^2} \ ,
\end{equation}
where $\rho_{\rm s}$ and $r_{\rm s}$ are characteristic density and
distance scales, respectively (see Navarro et al. 1997). These two
parameters are not independent, but related to a single parameter,
which can be taken as the cluster mass. It is important to note that
$\rho_{\rm s}$ and $r_{\rm s}$ also depend on the cosmological model,
hence the lensing properties of haloes with identical mass are
different in different cosmological models, a crucial property which
the simpler singular isothermal sphere models do not share. The NFW
halo falls off more steeply than isothermal at radii beyond $r_{\rm
s}$, but flattens towards the halo centre. These different features
lead to markedly different lensing properties of the NFW compared to
the SIS model, as will be shown below (see also the discussion in
Perrotta et al. 2002).

For axially symmetric lens models, the lens equation is essentially
one-dimensional. Define the optical axis as the line running from the
observer through the lens centre, and introduce physical distances
from the optical axis $\xi$ and $\eta$, respectively, on the lens and
source planes. Fixing a length scale $\xi_0$ on the lens plane, we can
define the dimensionless distance $x\equiv\xi/\xi_0$. The chosen
length scale $\xi_0$ is projected onto the length scale
$\eta_0=\xi_0\,D_{\rm s}/D_{\rm l}$ on the source plane, where $D_{\rm
l}$ and $D_{\rm s}$ are the angular-diameter distances to the lens and
source planes, respectively. In analogy to the lens plane, a
dimensionless distance from the optical axis $y\equiv\eta/\eta_0$ can
now be defined on the source plane. The lens equation, relating the
position of an image on the lens plane to that of its source on the
source plane is then
\begin{equation}
  y=x-\alpha(x) \ ,
\end{equation}
where $\alpha(x)$ is the reduced deflection angle at distance $x$ from
the lens centre, caused by the lensing mass distribution. It is the
gradient of the lensing potential $\psi$,
\begin{equation}
  \alpha(x)=\frac{\d\psi(x)}{\d x}\;.
\end{equation}
See, e.g.~Schneider, Ehlers \& Falco (1992) and Narayan \& Bartelmann
(1997) for more details.

Local imaging properties are described by the Jacobian matrix of the
lens mapping. It has two eigenvalues, $\lambda_{\rm r}$ and
$\lambda_{\rm t}$, which describe the image distortion in the radial
and the tangential directions, respectively. For an axially symmetric
lens, they are
\begin{equation}
  \lambda_{\rm r}(x)=1-\frac{\d\alpha}{\d x}\;,\quad
  \lambda_{\rm t}(x)=1-\frac{\alpha}{x} \ .
\end{equation}

Analytic expressions for the lensing potential $\psi(x)$ at any
distance $x$ are easily obtained for the SIS and the NFW lens models.
Therefore, the eigenvalues of the Jacobian matrix can
straightforwardly be computed for both lens models considered here.

For a SIS lens, the lensing potential is
\begin{equation}
  \psi(x)=|x|\;,
\label{eq:sisPsi}
\end{equation}
if $\xi_0=4\pi(\sigma_v/c)^2\,D_{\rm l}D_{\rm ls}/D_{\rm s}$ is chosen
as a length scale, where $D_{\rm ls}$ is the angular diameter distance
between lens and source. Hence, the deflection angle and the
eigenvalues are
\begin{equation}
  \alpha(x)=\frac{x}{|x|}\;,\quad
  \lambda_{\rm r}=1\;,\quad
  \lambda_{\rm t}=1-\frac{1}{|x|}\;.
\end{equation}

For an NFW lens, we take $\xi_0=r_{\rm s}$ for simplicity and define
$\kappa_{\rm s}\equiv\rho_{\rm s}r_{\rm s}\Sigma_{\rm cr}^{-1}$, where
$\Sigma_{\rm cr}=[c^2/(4\pi G)]\,[D_{\rm s}/(D_{\rm l}D_{\rm ls})]$ is
the critical surface mass density for strong lensing. We then have the
lensing potential
\begin{equation}
  \psi(x)=4\kappa_{\rm s}\left[\frac{1}{2}\ln^2\frac{x}{2}-
  2\,{\rm arctanh}^2\sqrt{\frac{1-x}{1+x}}\right]\;,
\label{eq:nfwPsi}
\end{equation}
which implies the deflection angle
\begin{equation}
  \alpha(x)=\frac{4\kappa_{\rm s}}{x}\,\left[
  \ln\frac{x}{2}+\frac{2}{\sqrt{1-x^2}}\,
  {\rm arctanh}\sqrt{\frac{1-x}{1+x}}
  \right]\;,
\end{equation}
from which the eigenvalues $\lambda_{\rm r,t}$ can straightforwardly
be derived. Several different aspects of lensing by haloes with NFW or
generalised NFW profiles can be found in Bartelmann (1996), Wright \&
Brainerd (2000), Li \& Ostriker (2002), Wyithe, Turner \& Spergel
(2001), Perrotta et al. (2002). It can easily be verified that the
potential [Eq.~(\ref{eq:nfwPsi})] satisfies the Poisson equation
$\nabla^2\psi=2\kappa$, with $\kappa$ given in Bartelmann (1996).

It is an important feature of the NFW lensing potential
[Eq.~(\ref{eq:nfwPsi})] that its radial profile is considerably less
curved near the centre than the SIS potential
[Eq.~(\ref{eq:sisPsi})]. Since the local imaging properties are
determined by the curvature of $\psi$, this immediately implies
substantial changes to the lensing properties. At fixed halo mass, the
critical curves of an NFW lens are closer to its centre than for a SIS
lens because of its flatter density profile. There, the potential is
less curved, thus the image magnification is larger and decreases more
slowly away from the critical curves. Therefore, NFW lenses are less
efficient in image splitting than SIS lenses, but comparably efficient
in image magnification. What is more important here is that any
additional shear added to a flat lensing potential (e.g.~by
asymmetries) much more strongly extends the critical curves than for a
steeper potential. We will see the consequences below.

The inverse of the eigenvalues $\lambda_{\rm r,t}$ of the Jacobian
matrix gives the radial and the tangential magnifications at the
radial distance $x$ from the lens centre. The points satisfying the
conditions $\lambda_{\rm r,t}=0$ form the radial or tangential
critical curves, respectively, where the corresponding magnifications
tend to infinity. For SIS lenses, $\lambda_{\rm r}$ is always unity,
hence the radial critical curve does not exist in this case, so the
images are not radially magnified. The tangential critical curve is
the circle $|x|=1$. For NFW lenses, the critical curves have to be
found numerically. In the limit of small $\kappa_{\rm s}$, when the
critical curves are close to the lens centre, we find to first order
in $x$
\begin{equation}
  \lambda_{\rm t}\approx
  2\exp\left(-\frac{1+\kappa_{\rm s}}{2\kappa_{\rm s}}\right)
  \;,\quad
  \lambda_{\rm r}\approx\frac{\lambda_{\rm t}}{2.718}\;.
\end{equation}

By means of the lens equation, the critical curves are mapped to the
corresponding caustic curves in the source plane. For the axially
symmetric lenses we are considering, the tangential caustic
degenerates to the point where the optical axis intercepts the source
plane.

\subsection{Elliptical sources}
\label{section:ellsrc}

The length-to-width ratio of an image depends on the local lens
properties and on the ellipticity of the sources. We derive and
compare results for both circular and elliptical sources. Since the
source galaxies are typically much smaller than the length scale on
which the lens properties change substantially, we can assume that the
length-to-width ratio of the images of a circular source is given by
the ratio of the two eigenvalues $q_{\rm l}\equiv\lambda_{\rm
r}/\lambda_{\rm t}$ at the image position.

An elegant model for introducing elliptical sources into analytic
lensing calculations was recently proposed by Keeton (2001). Following
his approach, the observed length-to-width ratio $L/W$ is a function
of three variables, which are the intrinsic axis ratio $q_{\rm s}=a/b$
and the intrinsic position angle $\theta$ of the source, and the
eigenvalue ratio $q_{\rm l}$ of the lens at the image position:
\begin{equation}
  L/W=\left[\frac{T+(T^2-4Q)^{1/2}}{T-(T^2-4Q)^{1/2}}\right]^{1/2}\;,
\end{equation}
where
\begin{eqnarray}
  T &=& q_{\rm l}^2+q_{\rm s}^2+(q_{\rm l}^2-1)
  (q_{\rm s}^2-1)\cos^2\theta\;,\\
  Q &=& q_{\rm l}^2q_{\rm s}^2\;.
\end{eqnarray}
	
The length-to-width ratio $L/W$ of each image can be related to the
source position $y$ by means of the lens equation. For sources with
fixed intrinsic shape and orientation lensed by axially symmetric
lenses, $L/W$ monotonically decreases as the source is moved from
$y=0$ (where the degenerate tangential caustic is located) away from
the lens centre. Thus the lensing cross section for arcs with $L/W \ge
(L/W)_{\rm min}$ from sources with intrinsic axis ratio $q_{\rm s}$
and position angle $\theta$ can be written as
\begin{equation}
  \sigma_{(L/W)_{\rm min}}(q_{\rm s},\theta)=\pi y_{\rm min}^2\;,
\end{equation}
where $y_{\rm min}$ is the distance from the optical axis where $L/W$
falls below $(L/W)_{\rm min}$.

\begin{figure}
{\centering \leavevmode
\psfig{file=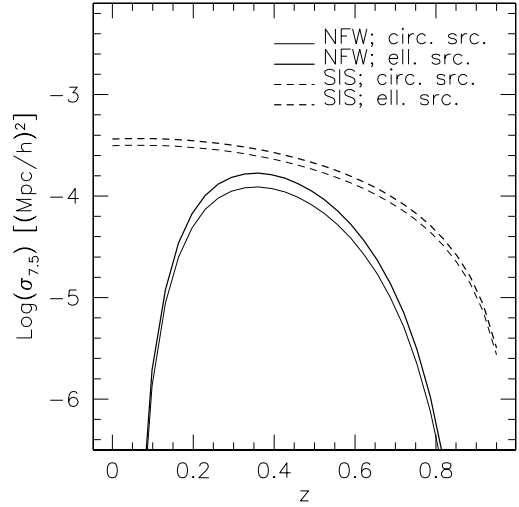,width=.40\textwidth}
}
\caption{Cross sections $\sigma_{7.5}$  for the formation of arcs with 
  length-to-width ratio larger than $7.5$ as functions of lens
  redshift. Results are shown for a virial lens mass of
  $M=10^{15}\,h^{-1}\,M_\odot$ in a $\Lambda$CDM cosmology. Two
  different lens models are used: the axially symmetric NFW lens model
  (solid curves), and the singular isothermal sphere for comparison
  (dashed lines). Thick and thin curves were obtained assuming
  intrinsically circular and elliptical sources, respectively.}
\label{figure:srcmodels}
\end{figure}

In order to compute the final lensing cross section, we have to
average these cross sections over all the possible values of $q_{\rm
s}$ and $\theta$. For doing so, we assume that $\theta$ and $q_{\rm
s}^{-1}$ are uniformly distributed in the ranges $[0,\pi]$ and
$[0.5,1]$, respectively. This agrees with the assumptions underlying
the numerical simulations. Therefore, the final lensing cross section
is given by
\begin{equation}
  \sigma_{(L/W)_{\rm min}}=\frac{4}{\pi}\int_{1}^{2}\int_{0}^{\pi/2}\,
  \sigma_{(L/W)_{\rm min}}(q_{\rm s},\theta)\,\d\theta
  \frac{\d q_{\rm s}}{q_{\rm s}^2}\;.
\end{equation}

\begin{figure*}
{\centering \leavevmode
\psfig{file=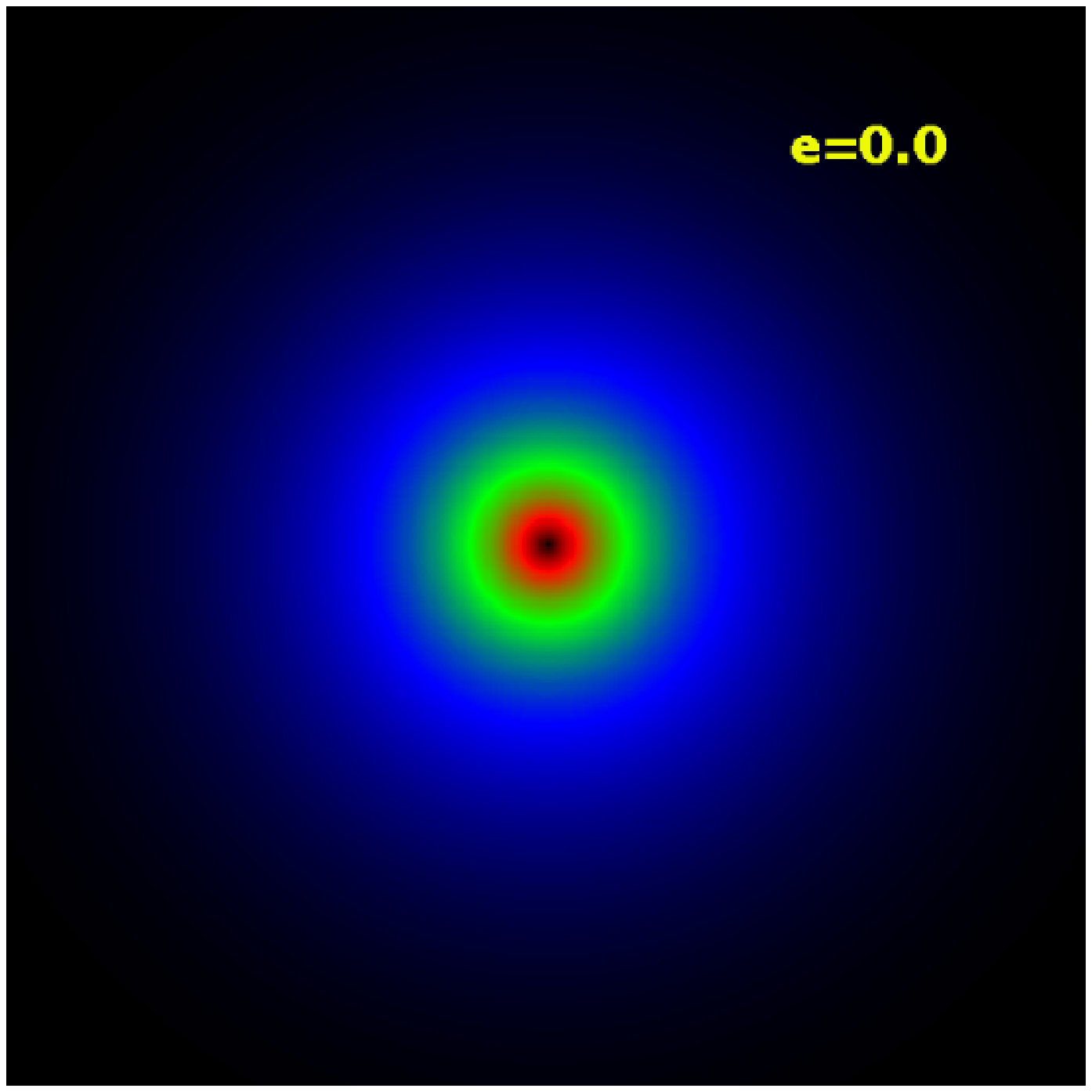,width=.30\textwidth} \hfil
\psfig{file=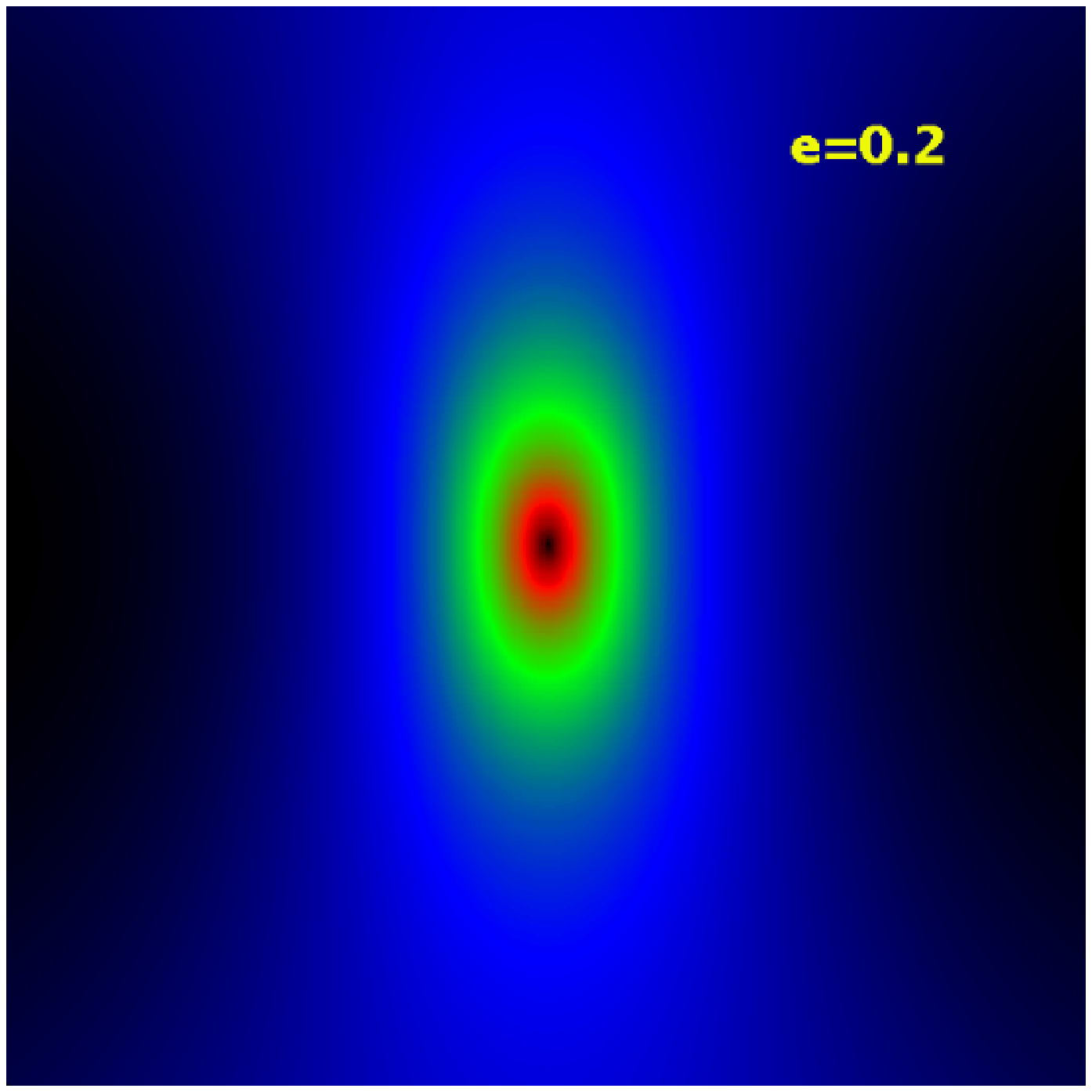,width=.30\textwidth} \hfil
\psfig{file=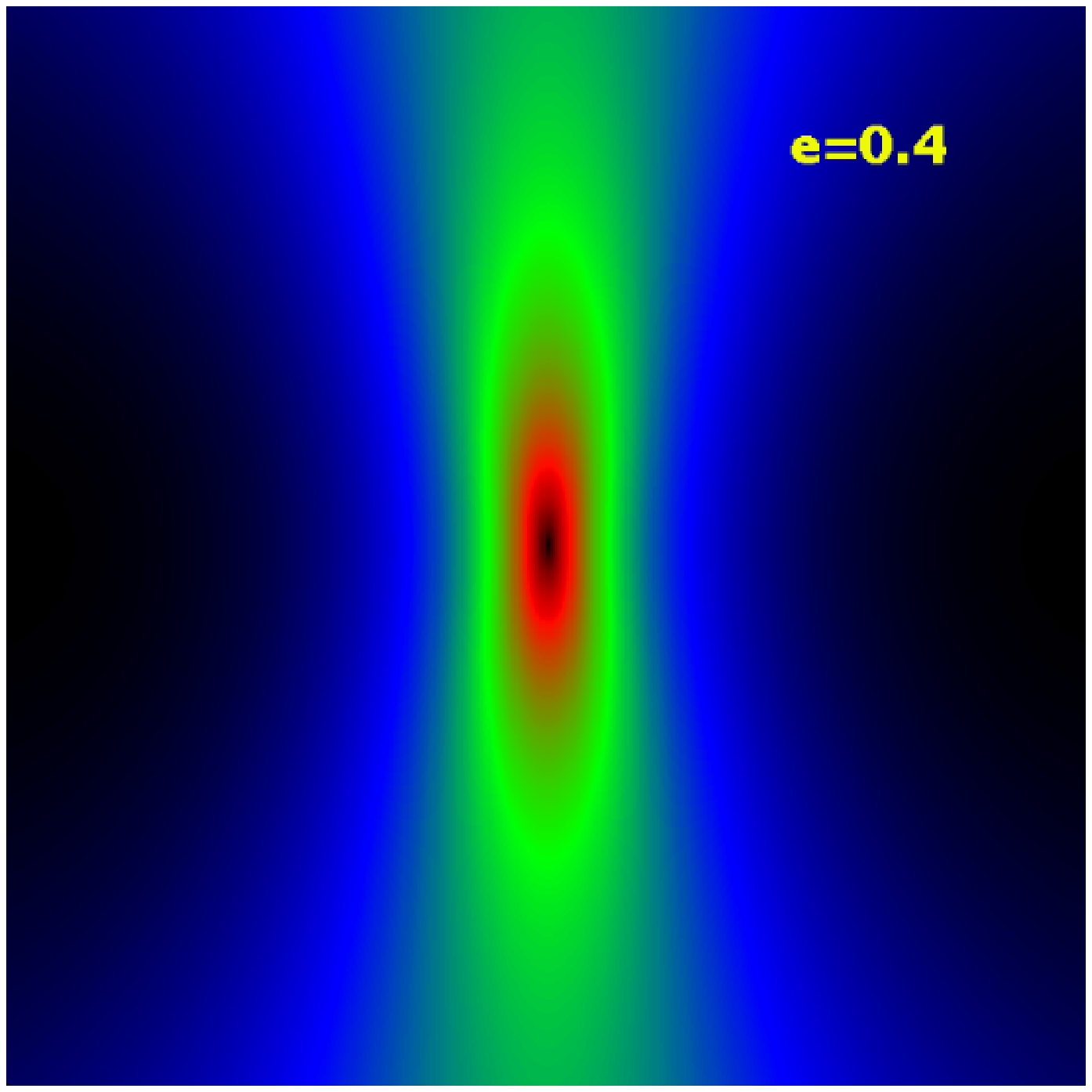,width=.30\textwidth}
}
\caption{Deflection-angle maps for the axially symmetric (left panel)
  and the elliptically distorted NFW lens model, for ellipticities
  $e=0.2$ (central panel) and $e=0.4$ (right panel). The side length
  of each panel is $\sim6'$. See the text for more detail.}
\label{figure:elldafields}
\end{figure*}

Both SIS and NFW lenses can produce multiple images for suitable
source positions. For SIS lenses, multiple images occur only for
sources lying inside the Einstein ring, which has the angular radius
$\theta_{\rm E}=(4\pi\sigma_v^2/c^2)\,(D_{\rm ls}/D_{\rm s})$. If this
is the case, the lens produces an arc and a counter-arc. As the source
moves towards the Einstein ring, the tangential magnification of the
counter-arc decreases and tends to zero. Then, the counter-arc
disappears (see Narayan \& Bartelmann 1997 for further detail). The
main ``arc'' remains, but its tangential magnification goes to unity
as the source moves away from the optical axis towards infinity.

For NFW lenses, multiple images occur only for sources with $y<y_{\rm
c,r}$, where $y_{\rm c,r}$ is the radius of the radial caustic. As
shown by Bartelmann (1996), the NFW density profile always has a
radial critical curve (and thus a radial caustic) for any combination
of $\rho_{\rm s}$ and $r_{\rm s}$. Sources inside the radial caustic
therefore always produce a radial image, an arc, and a counter-arc. We
neglect the radial image because it is radially oriented. As for SIS
lenses, the arc and the counter-arc have different length-to-width
ratios $L/W$. In fact, the counter-arc is radially magnified as the
source moves towards the radial caustic, and its length-to-width ratio
decreases more rapidly than that of the main tangential
arc. Therefore, two separate cross sections must be computed for both
SIS and NFW lens models, one for the arc and one for the counter-arc,
and the total lensing cross section is the sum of these two
contributions.

To give an example, Fig.~\ref{figure:srcmodels} shows the lensing
cross section for arcs with length-to-width ratio $L/W>7.5$ produced
from sources at redshift $z_{\rm s}=1$ by a lens with mass
$10^{15}\,h^{-1}\,M_\odot$ at redshifts between zero and unity.
Results are given for the $\Lambda$CDM model. The solid lines
correspond to NFW lenses, the dashed lines to SIS lenses. Thick and
thin lines indicate the results for elliptical and circular sources,
respectively. We use here the length-to-width ratio threshold of $7.5$
instead of $10$ in order to improve the statistics of numerically
simulated arcs, and to reduce corresponding fluctuations in the cross
sections.

As expected, the lensing cross section of the NFW model is smaller
than for SIS lenses because of its flatter density profile. In
particular, NFW lenses lose their strong-lensing efficiency when the
lens approaches the observer or the sources. Conversely, due to its
unrealistically steep and scale-free density profile, the SIS remains
an efficient strong lens even when it is located very close to the
observer or to the source.

The relative increase of the cross section for the intrinsically
elliptical sources is virtually independent of the lens redshift, but
it does depend on the lens model. In fact, the relative change in the
lensing cross sections for NFW lenses is approximately twice as high
as for SIS lenses, quite independent of the lens redshift. This is
again due to the shallower lensing potential of NFW compared to SIS
lenses.

\subsection{Elliptical models}
\label{section:ellcross}

The construction of lens models with elliptical or pseudo-elliptical
isodensity contours is generally quite complicated (see e.g. Kassiola
\& Kovner 1993; Kormann, Schneider \& Bartelmann 1994; Golse \& Kneib
2002). Starting from an elliptical lensing potential is
computationally much more tractable, but has the disadvantage that the
mass distribution corresponding to the elliptical potential can become
dumbbell-shaped even for moderate ellipticities, which is unwanted for
galaxy lenses. Galaxy clusters, however, are less relaxed and exhibit
substructure, so for them dumbbell-shaped mass distributions are
uncritical. For this reason, we prefer constructing elliptical cluster
lens models starting directly from the effective lensing potential.
Moreover, since the NFW density profile gives a much more realistic
reproduction of the simulated clusters than the SIS, we chose to
generalise only the NFW lens model to the elliptical case.

The lensing potential of an axially symmetric NFW lens was given in
Eq.~(\ref{eq:nfwPsi}) above. We introduce the ellipticity
$e\equiv1-b/a$, where $a$ and $b$ are the major and minor axes of the
ellipse, by substituting
\begin{equation}
  x\rightarrow{\cal X}=\sqrt{\frac{x_1^2}{(1-e)}+x_2^2(1-e)}\;,
\end{equation}
where $x_1$ and $x_2$ are the two Cartesian components of $x$,
$x^2=x_1^2+x_2^2$. This ensures that the mass inside circles of fixed
radius remains constant as the ellipticity changes.

The Cartesian components of the deflection angle are then
\begin{eqnarray}
  \alpha_1 &=& \frac{\partial\psi}{\partial x_1}=
  \frac{x_1}{(1-e){\cal X}}\hat{\alpha}({\cal X})\;,\nonumber\\
  \alpha_2 &=& \frac{\partial\psi}{\partial x_2}=
  \frac{x_2(1-e)}{\cal X}\hat{\alpha}({\cal X})\;,
\label{equation:angles}
\end{eqnarray}
where $\hat{\alpha}({\cal X})$ is the unperturbed
(i.e.~axially-symmetric) deflection angle at the distance ${\cal X}$
from the lens centre.

Using these formulae, deflection-angle fields for different values of
the ellipticity $e$ are readily computed. Some examples for
deflection-angle maps are displayed in
Fig.~\ref{figure:elldafields}. Obviously, the shape of the contours
becomes more and more elliptical as $e$ is increased. We analyse the
lensing properties of these deflection-angle fields using the
ray-tracing technique and, finally, compute the lensing cross
sections, adopting the same techniques as applied to the numerical
models.

\begin{figure*}
{\centering \leavevmode
\psfig{file=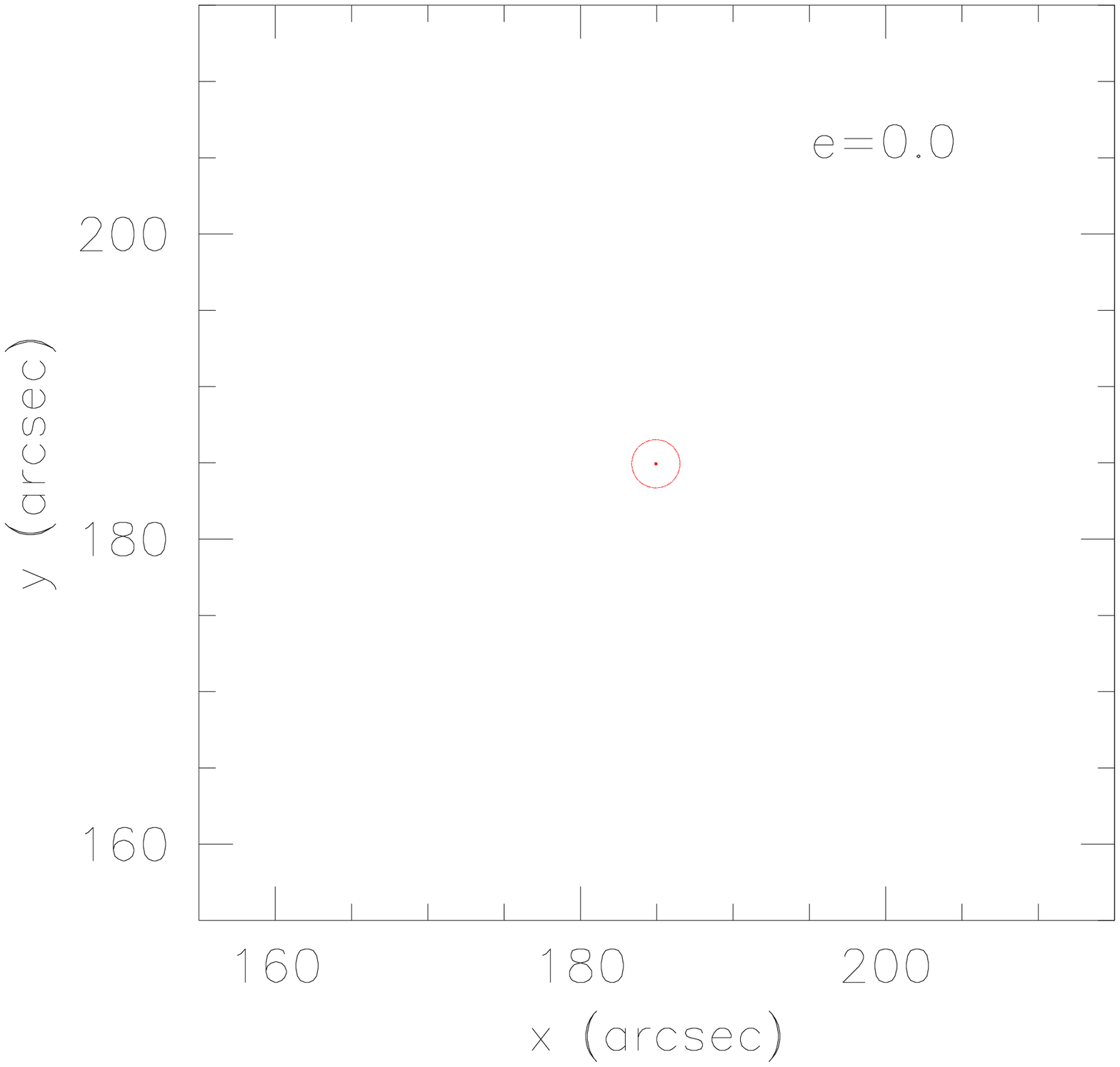,width=.30\textwidth} \hfil
\psfig{file=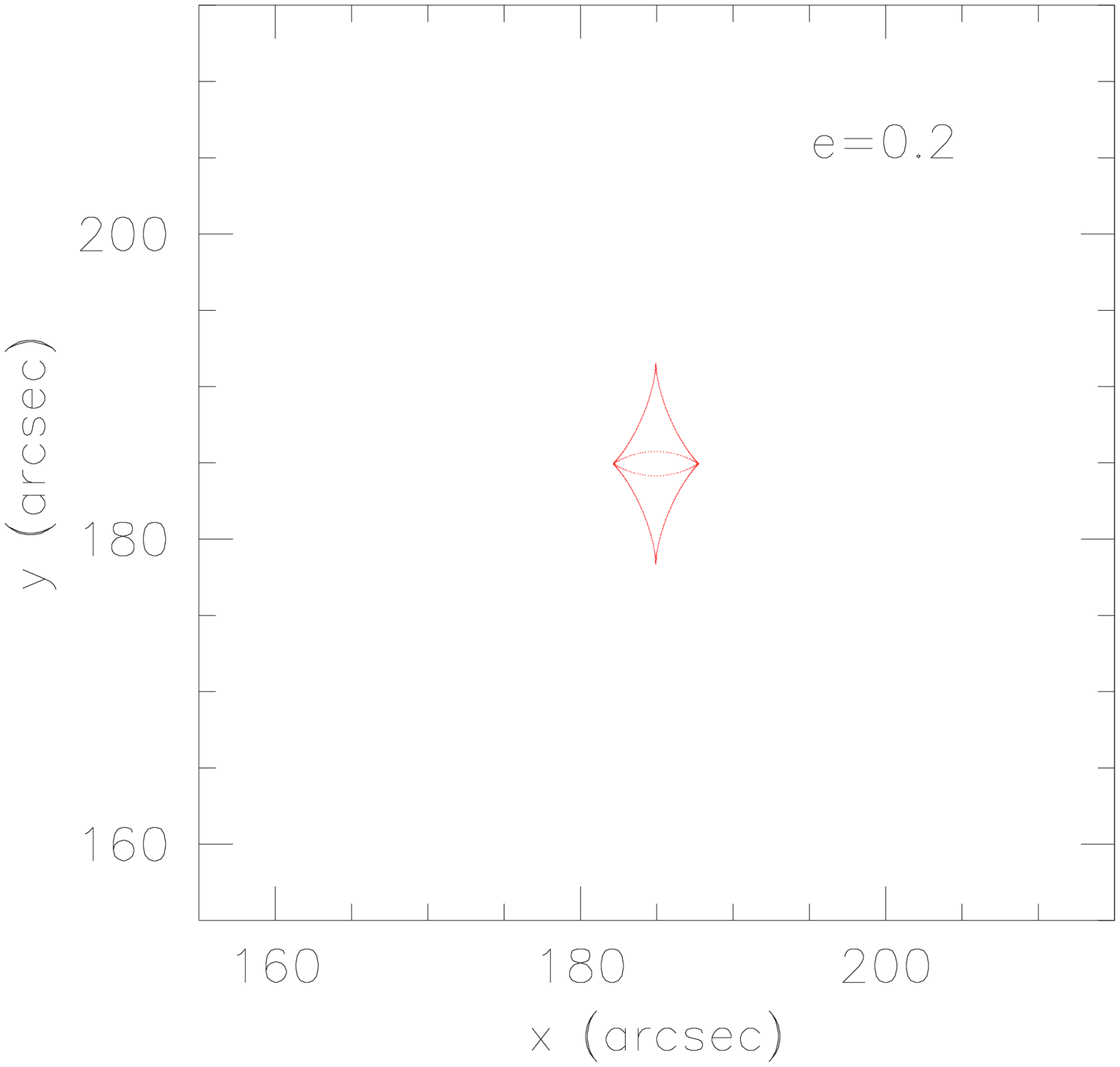,width=.30\textwidth} \hfil
\psfig{file=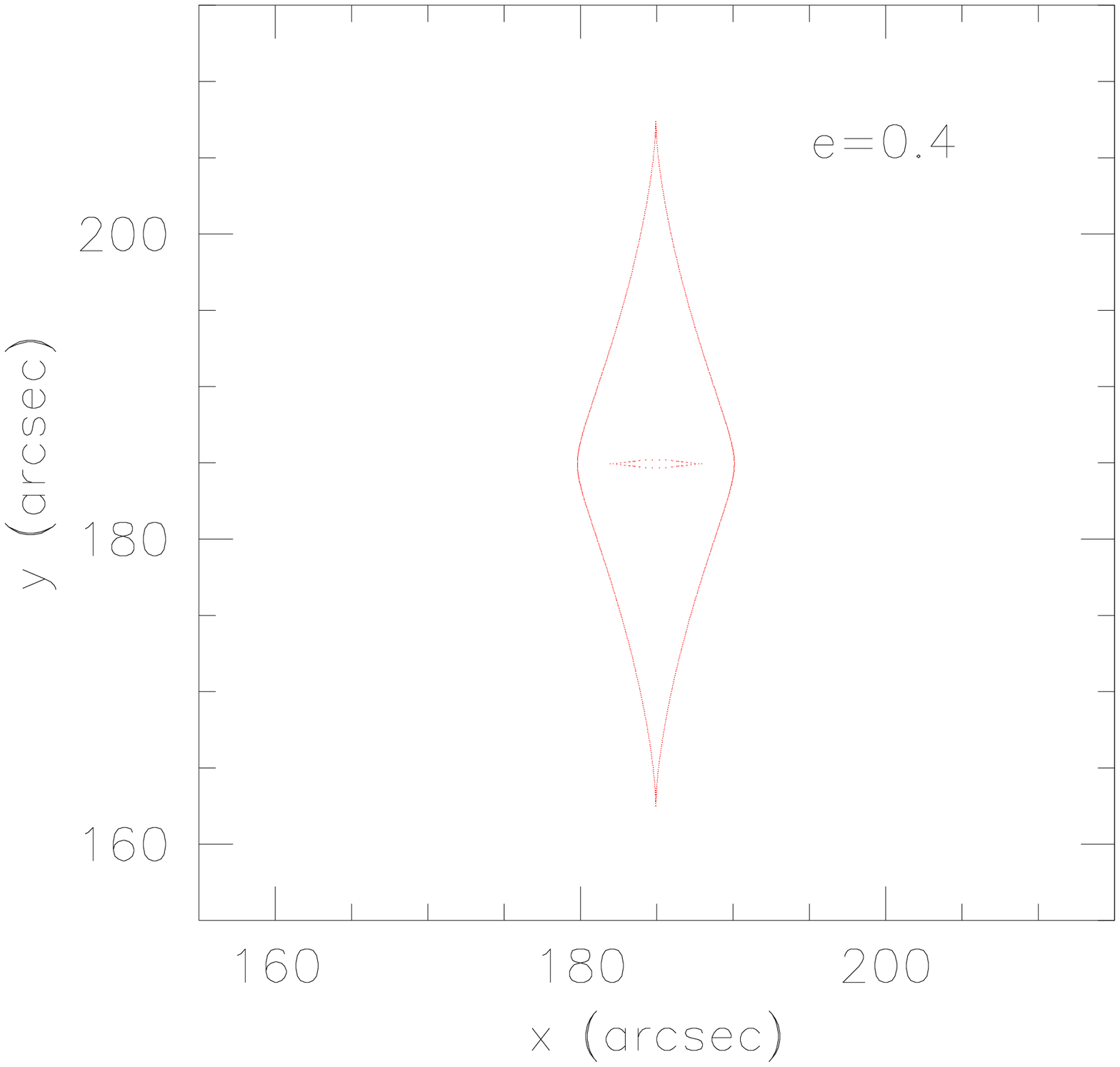,width=.30\textwidth}
}
\caption{Caustic curves produced by an NFW halo with mass
  $M=10^{15}\,h^{-1}\,M_\odot$ at redshift $z=0.3$ in a $\Lambda$CDM
  universe. Different panels show results for different ellipticities
  $e$ of the lensing potential: $e=0$, i.e.~axially symmetric (left
  panel); $e=0.2$ (central panel) and $e=0.4$ (right panel). The
  figure demonstrates that even small or moderate elliptical
  distortions substantially stretch the caustic curves, and thus cause
  the arc cross sections to grow considerably.}
\label{figure:caustics}
\end{figure*}

\begin{figure}
{\centering \leavevmode
\psfig{file=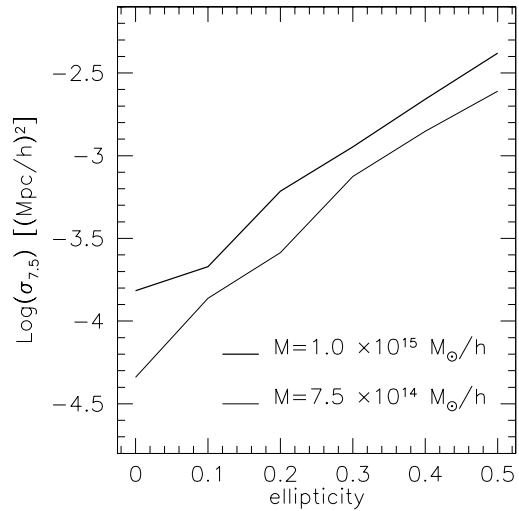,width=.40\textwidth}
}
\caption{Strong-lensing cross section $\sigma_{7.5}$ of elliptically
  distorted NFW lenses for arcs with length-to-width ratio larger than
  $7.5$ as a function of the ellipticity $e$ of the lensing
  potential. Tick and thin lines were calculated for haloes of mass
  $M=10^{15}\,h^{-1}\,M_\odot$ and
  $M=7.5\,\times10^{14}\,h^{-1}\,M_\odot$, respectively, both placed
  at $z=0.3$ in a $\Lambda$CDM cosmology.}
\label{figure:ellgrow}
\end{figure}

Increasing the ellipticity of the lensing potential strengthens the
shear field of the lens, and consequently the tangential caustic
expands and changes. Examples for the change of the caustics with
ellipticity $e$ are shown in Fig.~\ref{figure:caustics}, which refers
to a halo with mass $M=10^{15}\,h^{-1}\,M_{\odot}$ at redshift
$z=0.3$, and the underlying cosmology is the $\Lambda$CDM model. As
discussed above, the radial and tangential caustics are a circle and a
point, respectively, for the axially symmetric models. Increasing $e$,
the caustics stretch, develop cusps, and enclose a growing area. Thus
the lensing cross sections are expected to grow rapidly. This is
confirmed by Fig.~\ref{figure:ellgrow}, which shows cross sections for
arcs with $L/W\ge7.5$. These results were obtained through ray-tracing
simulations in which deflection angle maps for two haloes of mass
$M=7.5\times10^{14}\,h^{-1}\,M_{\odot}$ and
$M=10^{15}\,h^{-1}\,M_{\odot}$ were used. Again, we adopt the
$\Lambda$CDM model universe and put the lens at redshift $z=0.3$.
Increasing the ellipticity of the lensing potential from $e=0$ to
$e=0.5$, the cross section increases approximately by a factor of
$30$, almost independently of the lens mass considered.

\section{Comparison of Analytic and Numerical models}
\label{section:compa}

We can now compare the strong lensing cross sections of the numerical
and analytic models introduced in the preceding sections. For that
purpose, we focus on the lensing cross sections for the formation of
arcs with length-to-width ratios $L/W$ larger than $7.5$ and $10$. As
described before, we compare them to the lensing properties of five
numerical models of galaxy cluster haloes, picking simulation
snapshots at twelve different redshifts between zero and unity. For
clarity, we present here the results obtained for the most massive
halo only. The behaviour of the cross sections for the other numerical
models is in good qualitative and quantitative agreement with that
obtained for this cluster.

\begin{figure*}
{\centering \leavevmode
\psfig{file=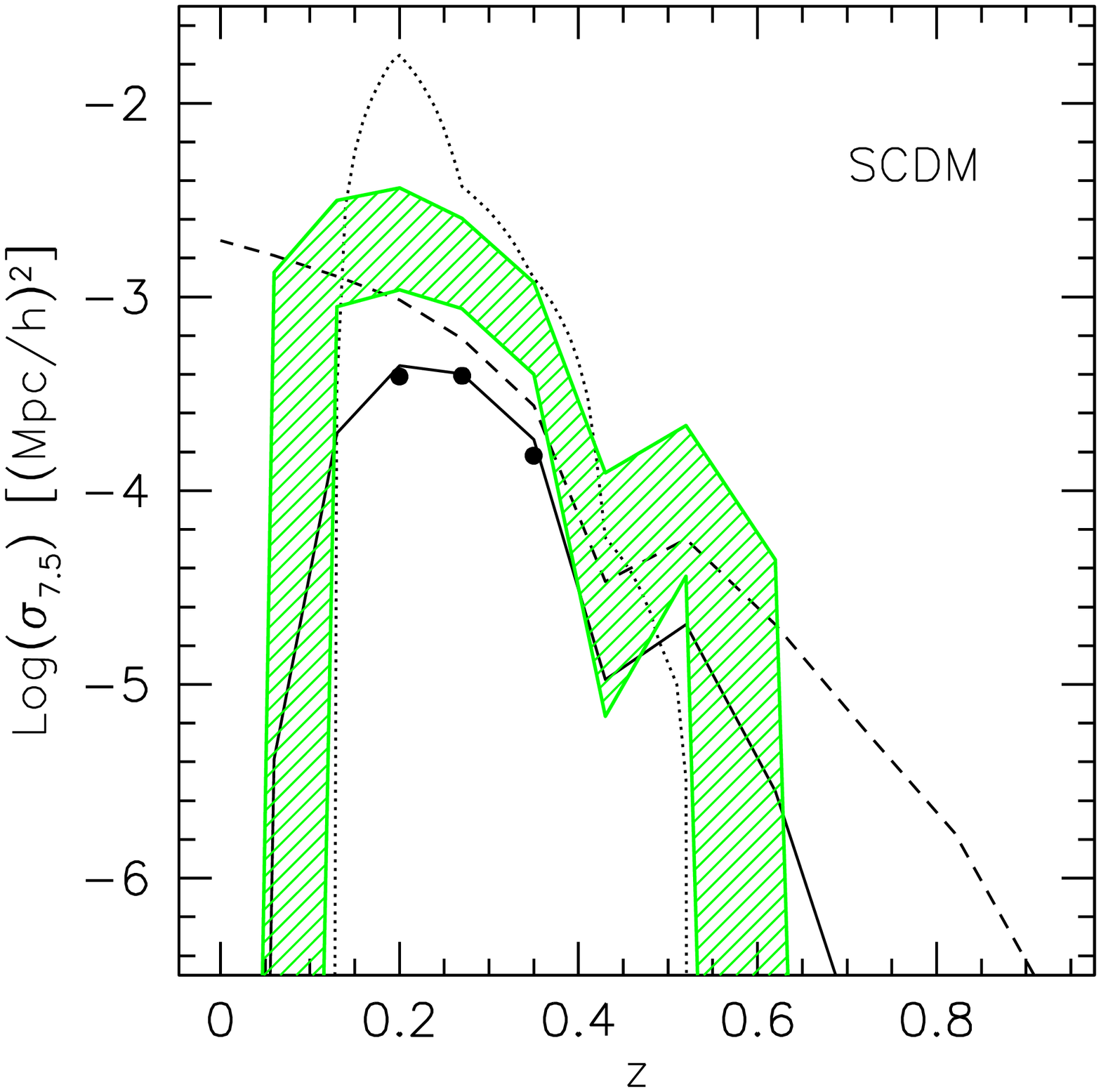,width=.30\textwidth} \hfil
\psfig{file=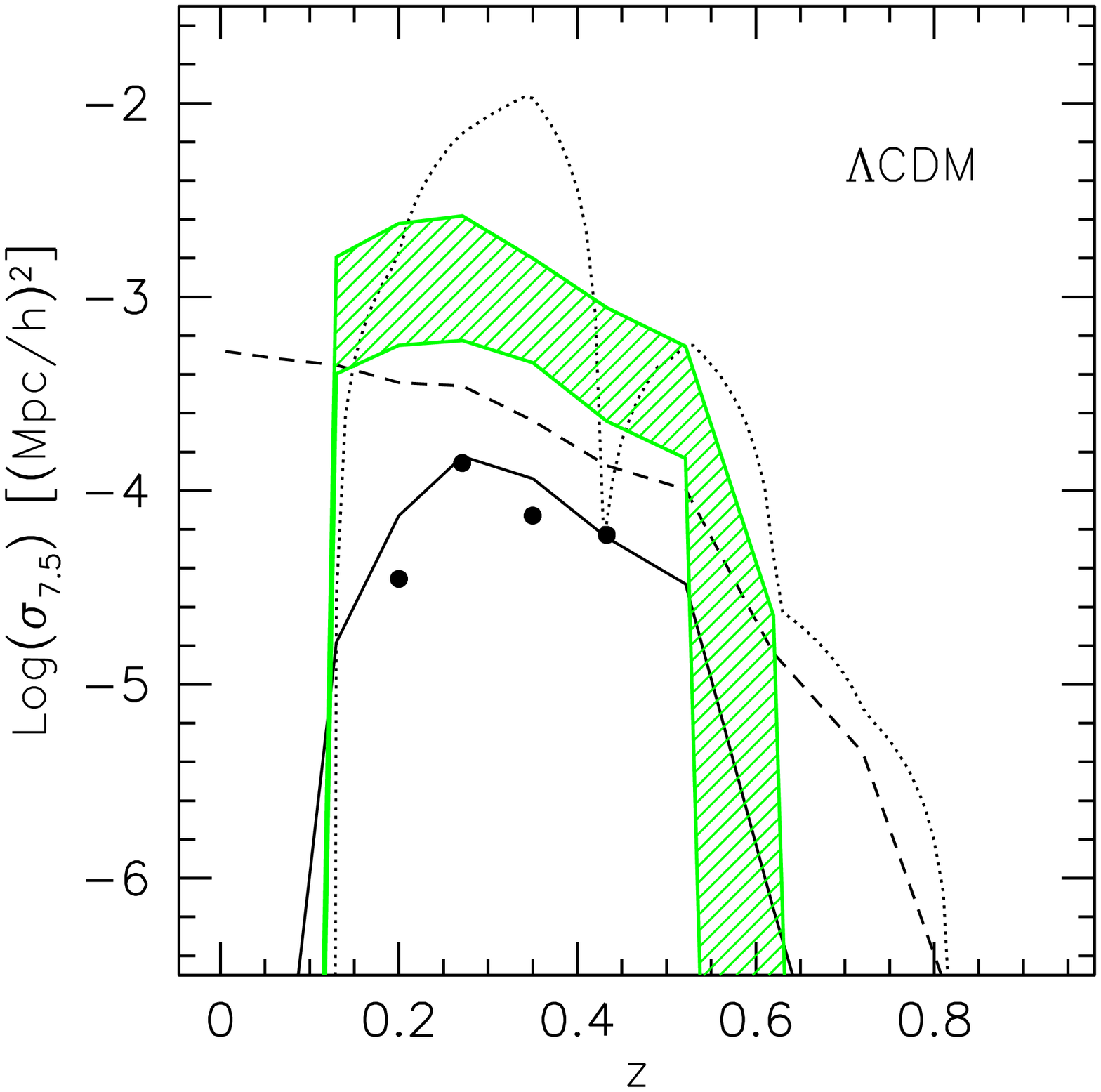,width=.30\textwidth} \hfil
\psfig{file=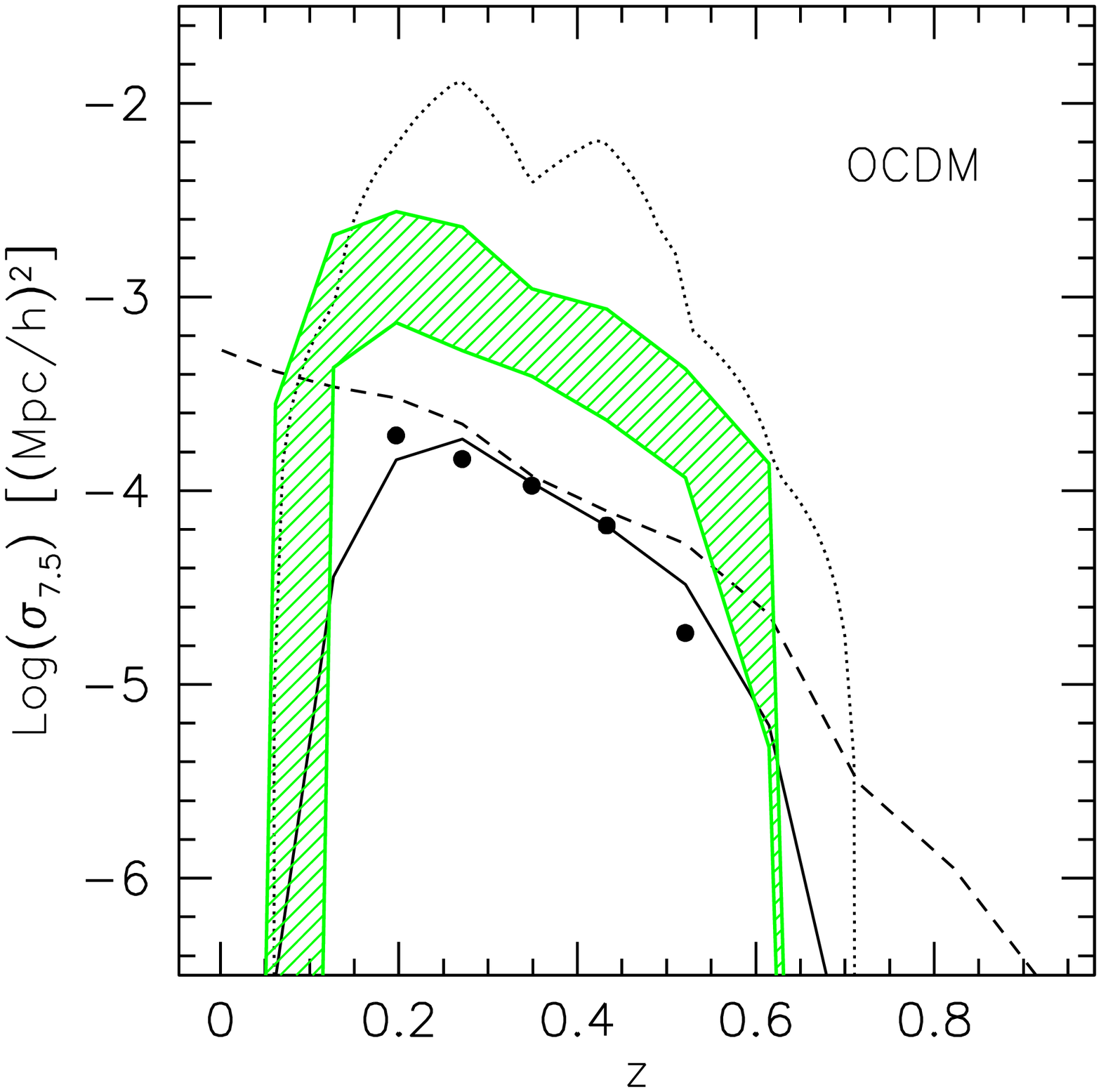,width=.30\textwidth} \hfil
\psfig{file=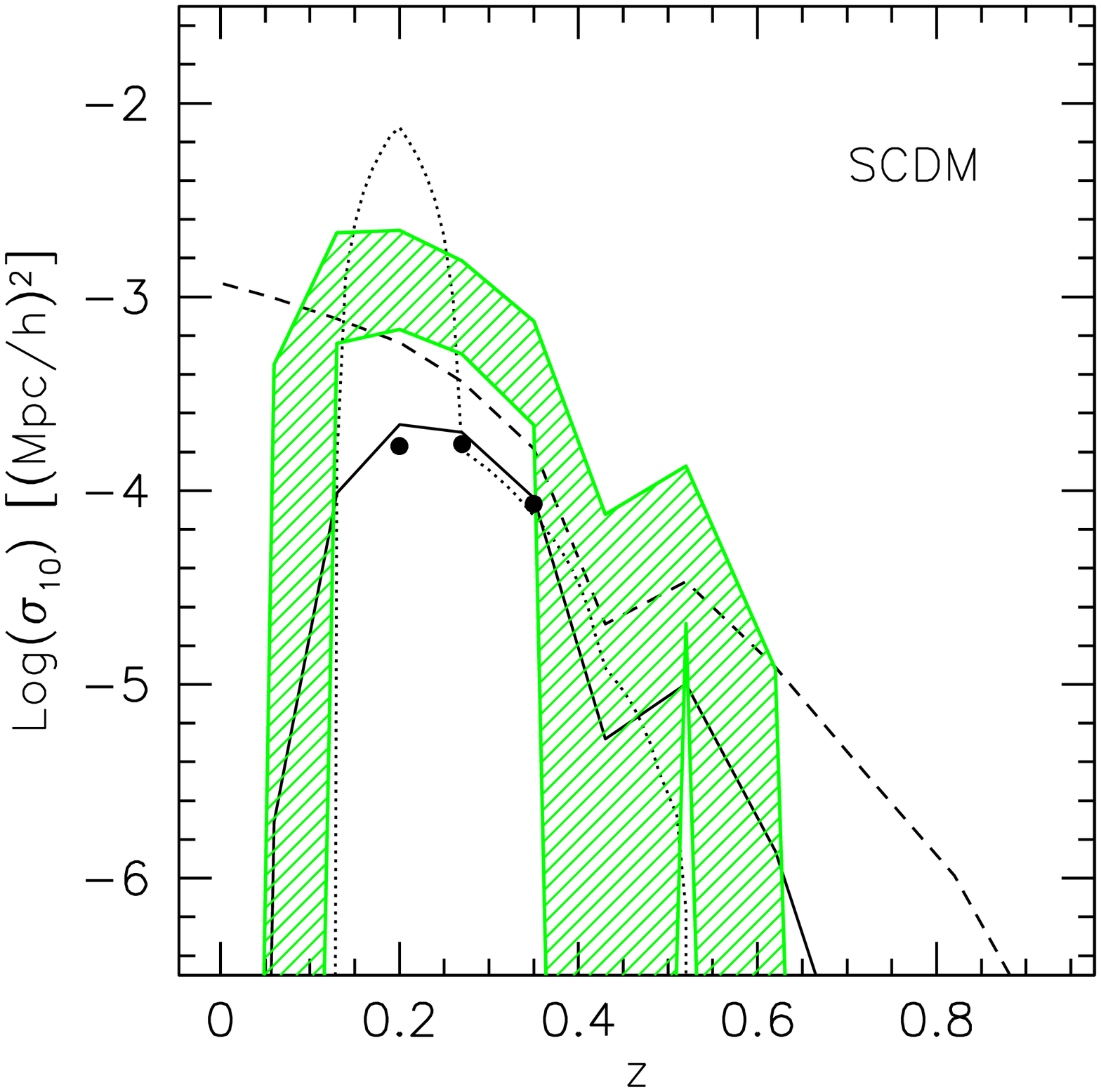,width=.30\textwidth} \hfil
\psfig{file=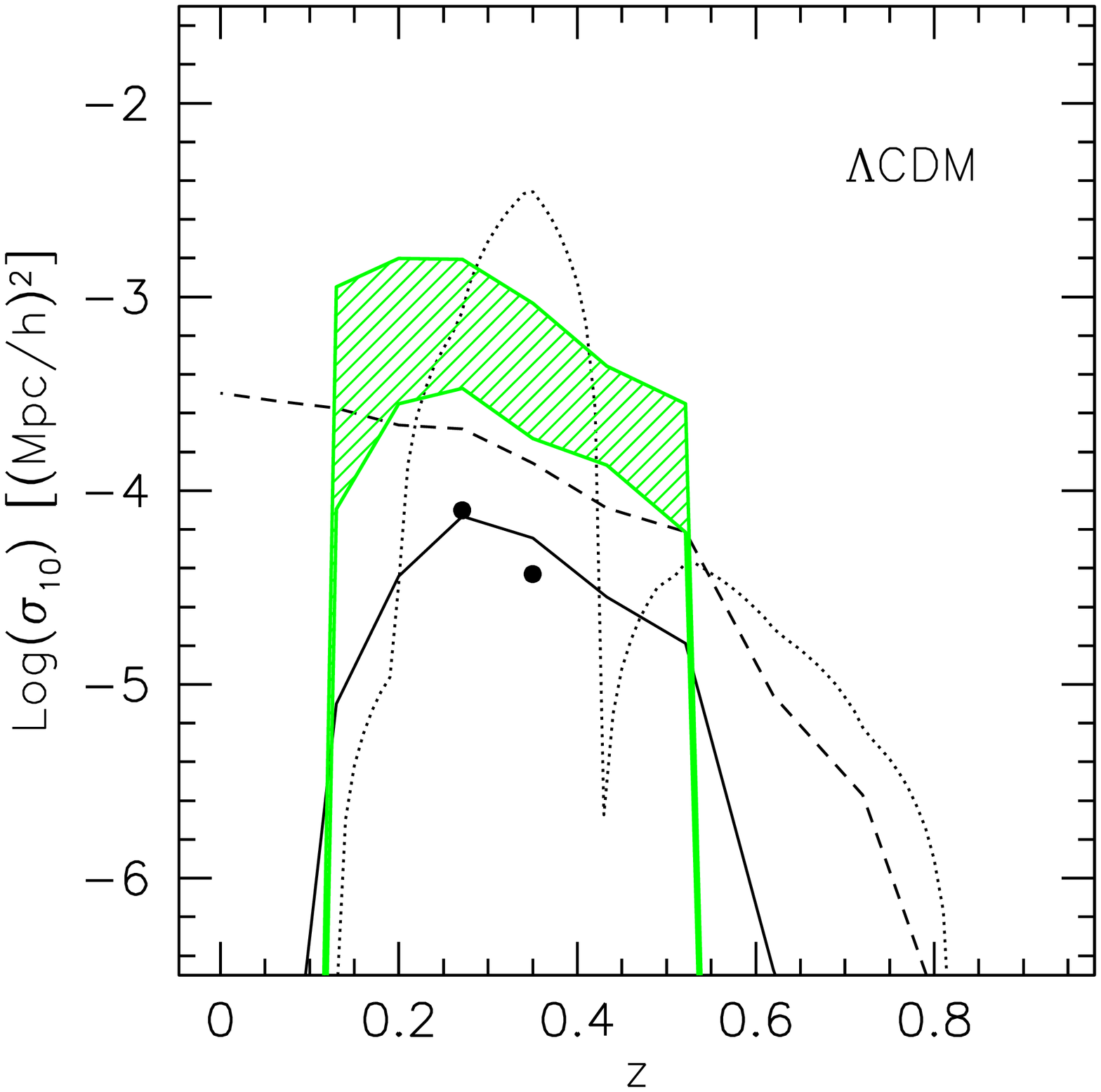,width=.30\textwidth} \hfil
\psfig{file=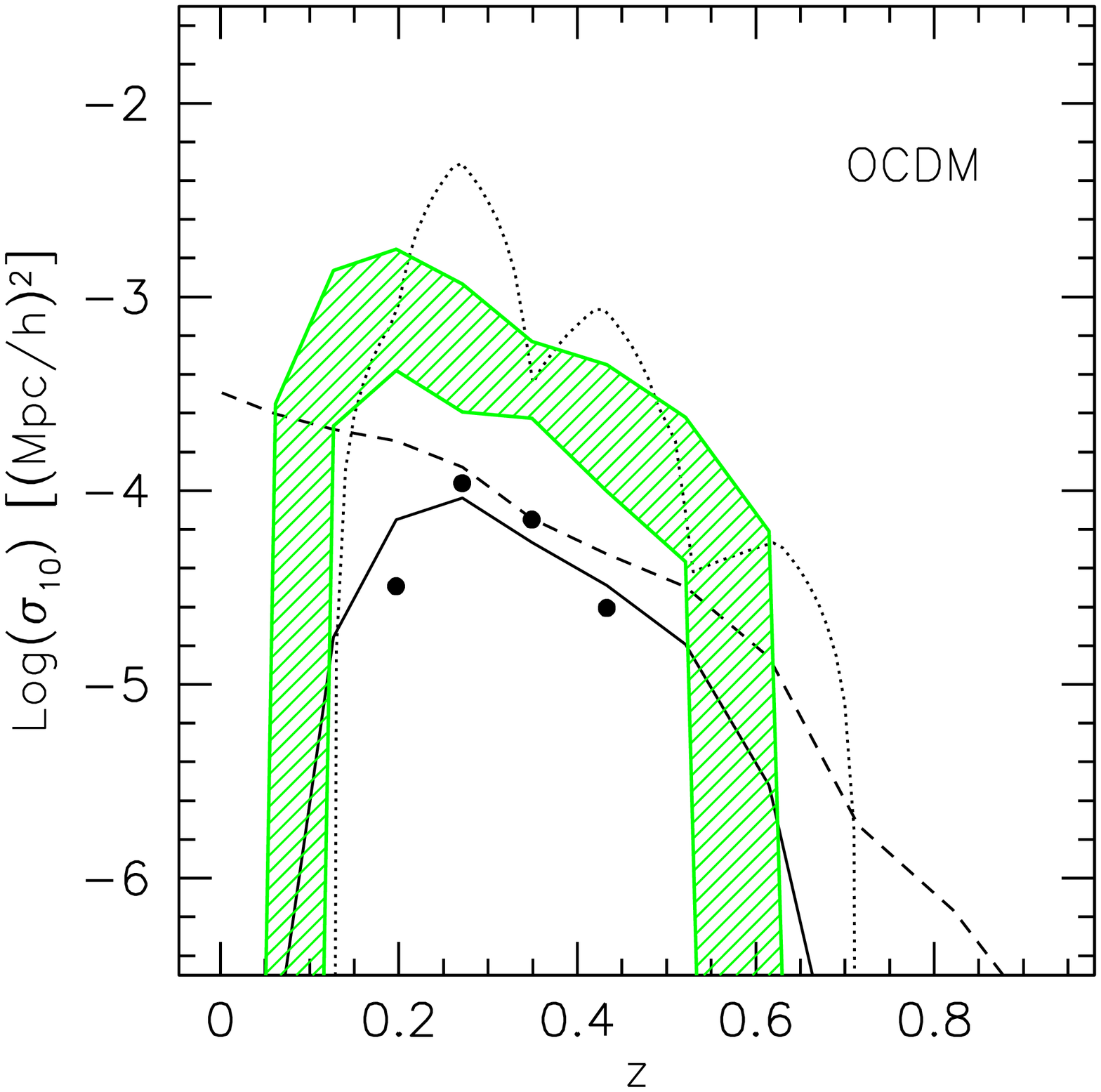,width=.30\textwidth}
}
\caption{Lensing cross sections for arcs with length-to-width ratio
  larger than $7.5$ ($\sigma_{7.5}$; top panels) and $10$
  ($\sigma_{10}$; bottom panels) as a function of the lens
  redshift. Results for different cosmologies are shown: SCDM (left
  panels), $\Lambda$CDM (central panels) and OCDM (right
  panels). Dotted lines show results obtained from the analysis of the
  most massive halo in the numerically simulated sample; solid and
  dashed lines represent the cross sections predicted by the axially
  symmetric NFW and SIS models, respectively, with the same virial
  mass as the numerically modelled halo. The shaded regions mark the
  cross sections obtained with the elliptically distorted NFW lens
  model, with ellipticities ranging between $e=0.2$ and $e=0.4$ (lower
  and upper limits, respectively). The filled dots are the lensing
  cross sections found by ray-tracing simulations with the deflection
  angle maps of the axially symmetric NFW lens model. Their agreement
  with the analytic curves demonstrates the reliability of our
  numerical method.}
\label{figure:comparison}
\end{figure*}

The results are illustrated in Fig.~\ref{figure:comparison}, where the
dotted lines refer to the fully numerically simulated cluster, while
solid and dashed lines represent the cross sections of NFW and SIS
lenses, respectively, having the same virial mass as the numerical
cluster model. Finally, the shaded regions in the same plots show the
cross sections obtained by elliptically distorting the NFW lensing
potential with ellipticities $e$ in the range between $e=0.2$ and
$e=0.4$ (lower and upper limits, respectively). Results are shown for
$L/W=7.5$ (upper panels) and $L/W=10$ (lower panels), and they were
obtained for the SCDM (left panels), $\Lambda$CDM (central panels),
and OCDM models (bottom panel).

First, we checked the calibration of the analytic relative to the
numerical cross sections. For doing so, we performed ray-tracing
simulations using the axially symmetric deflection angle maps for the
NFW lens model, and used them for determining the strong-lensing cross
sections. The results are shown as filled dots in the panels of
Fig.~\ref{figure:comparison}. The very good agreement with the
analytic estimates obtained as explained in
Sect.~\ref{section:sphcross} demonstrates the reliability of our
numerical technique.

The general trends in the lensing cross sections shown in
Fig.~\ref{figure:comparison} can be understood as follows. The
strong-lensing efficiency of a mass distribution depends on several
factors. First, for the light coming from the sources to be focused on
the observer, the lens must be located at a suitable distance from
both the observer and the sources. Second, the larger the (virial)
mass of the lens is, the stronger are the lensing effects it produces
close to its centre. Finally, the more concentrated the lens is, the
thinner are the long arcs expected to be.

Since the lens mass grows with decreasing redshift because new
material is accreted by the halo and deepens its potential well, the
lensing cross sections are expected to grow as well. On the other
hand, when the lens is too close to the observer, the cross section is
geometrically suppressed, unless the lens surface density profile is
sufficiently steep and scale-free, as for the SIS lenses. In fact, in
this case the focusing by the lens is strong enough to allow observers
to see strongly distorted images of background sources also in very
near lenses, i.e.~at relatively small redshifts (see also
Fig.~\ref{figure:srcmodels}).

Moreover, the results are expected to depend on cosmology. In fact, in
the OCDM and $\Lambda$CDM models, galaxy clusters become efficient
lenses at higher redshift than in the SCDM scenario, because they form
earlier and are centrally more concentrated. This is confirmed by our
numerical results. In the Einstein-de Sitter case both $\sigma_{7.5}$
and $\sigma_{10}$ are completely negligible for redshifts $z\ga
0.4$--$0.5$ and peak at $z\approx0.2$, while for low-density models
the cross sections are largest in the redshift range $0.3\la
z\la0.5$. A more complete discussion on the dependence of lensing
cross sections on cosmology can be found in Bartelmann et al. (1998).

In this paper, we concentrate on comparing the cross sections of the
analytic and numerical models. As Fig.~\ref{figure:comparison} shows,
the numerical models generally have much larger cross sections than
the analytic models. In particular, the cross sections for axially
symmetric NFW lenses are almost two orders of magnitude smaller, quite
independent of the cosmological models considered. For SIS lenses, the
discrepancy with the numerical models is only partially compensated by
the unrealistically steep central density profile, but the estimated
values of $\sigma_{7.5}$ and $\sigma_{10}$ remain too low. Introducing
the elliptical distortion into the NFW lens model allows the cross
section to increase by roughly an order of magnitude compared to the
axially symmetric NFW model, but even then the analytic cross sections
fail to reproduce the numerical cross sections unless unrealistically
high values of $e$ are adopted.

Fig.~\ref{figure:comparison} also demonstrates that apart from the
cross-section amplitude, the analytic models miss another important
feature of the numerical results, which show steep increases and
decreases reflecting merger events. While a merging sublump approaches
a cluster, the cross section tends to increase because of the
increasing tidal field, and as the cluster relaxes following the
merger event, the cross section decreases again. It becomes quite
clear from Fig.~\ref{figure:comparison} that such events play an
important role in understanding realistic cluster cross sections, and
they cannot reasonably be captured in analytic models unless their
prime advantage of being computationally fast is sacrificed.

\subsection{Ellipticity estimates}

It is important to check which ellipticities are typical for the
lensing potential of the numerical clusters. To this end, we compare
the deflection-angle maps constructed for the numerical haloes with
those obtained for elliptical NFW lenses with identical virial
mass. We estimate the best-fit ellipticity $e$ by minimising the
mean-square deviation
\begin{equation}
  \chi^2=\sum_i\left(\frac{\alpha_i-\hat{\alpha}_i}
  {\hat{\alpha}_i}\right)^2\;,
\end{equation}
where $\alpha_i$ and $\hat{\alpha}_i$ are the deflection angles of the
light ray passing through the $i$-th grid point on the numerical and
the elliptical lens model, respectively. The summation is done over
all grid points contained in the central region of roughly
$\sim200\,h^{-1}\,{\rm kpc}$ comoving side length. This way, we
measure the ellipticity of the lensing potential in the central region
close to the critical curves which is the most relevant for our
purposes. For this test, we use the simulation snapshots at $z=0.27$,
which is approximately the redshift where the numerical haloes reach
their maximum lensing efficiency. The complete sample of five
numerical clusters and three independent projections per cluster was
used for this analysis.

The results are reported in Tab.~\ref{table:chitest}. Median
ellipticities of $e\sim0.3$ are found, with only a weak dependence on
cosmology. Indeed, only little evolution of the ellipticity is found
going from low to high-density cosmological models. The median
ellipticities in the OCDM and the $\Lambda$CDM models are slightly
smaller than in the SCDM case, as expected, given that clusters form
earlier in these models and have more time to relax. Anyway, as the
semi-interquartile ranges (SIQR) show, the ellipticity distributions
are quite broad, in particular for SCDM. The same analysis has also
been repeated using other simulation snapshots, corresponding to
redshifts in the range $0.2\la z\la0.7$. The results are quite similar
to the previous ones and therefore not given in the table. The typical
median of $e$ is around $0.3$, and exceeds $0.4$ in only a few cases.
In conclusion, the ellipticities measured in the lensing potential of
the numerical clusters are insufficient for reconciling the cross
sections of elliptically distorted NFW models with the fully numerical
results.

\begin{table}
\caption{Ellipticities $e$ of the lensing potential obtained by
  fitting the deflection-angle maps of the elliptically distorted NFW
  lens model to fully numerical deflection angles. We give the medians
  (second column) and semi-interquartiles ranges (SIQR; third column)
  for the ellipticities found in our cluster sample for different
  cosmological models (indicated in the first column). Results are
  shown for the simulation snapshots at redshift $z=0.27$.}
\begin{center}
\begin{tabular}{lcc}
\hline\hline
  cosmological model & median & SIQR \\
\hline\hline
  SCDM         &  0.320 & 0.102 \\
  $\Lambda$CDM &  0.305 & 0.032 \\
  OCDM         &  0.270 & 0.052 \\
\hline\hline
\end{tabular}
\end{center}
\label{table:chitest}
\end{table}

\subsection{Substructure estimates}

The remaning difference between numerical and analytical cross
sections must be attributed to some factors which are missing from the
analytical models. The most important of those is certainly the
presence of substructure in the lensing mass distribution.

\begin{figure*}
{\centering \leavevmode
\psfig{file=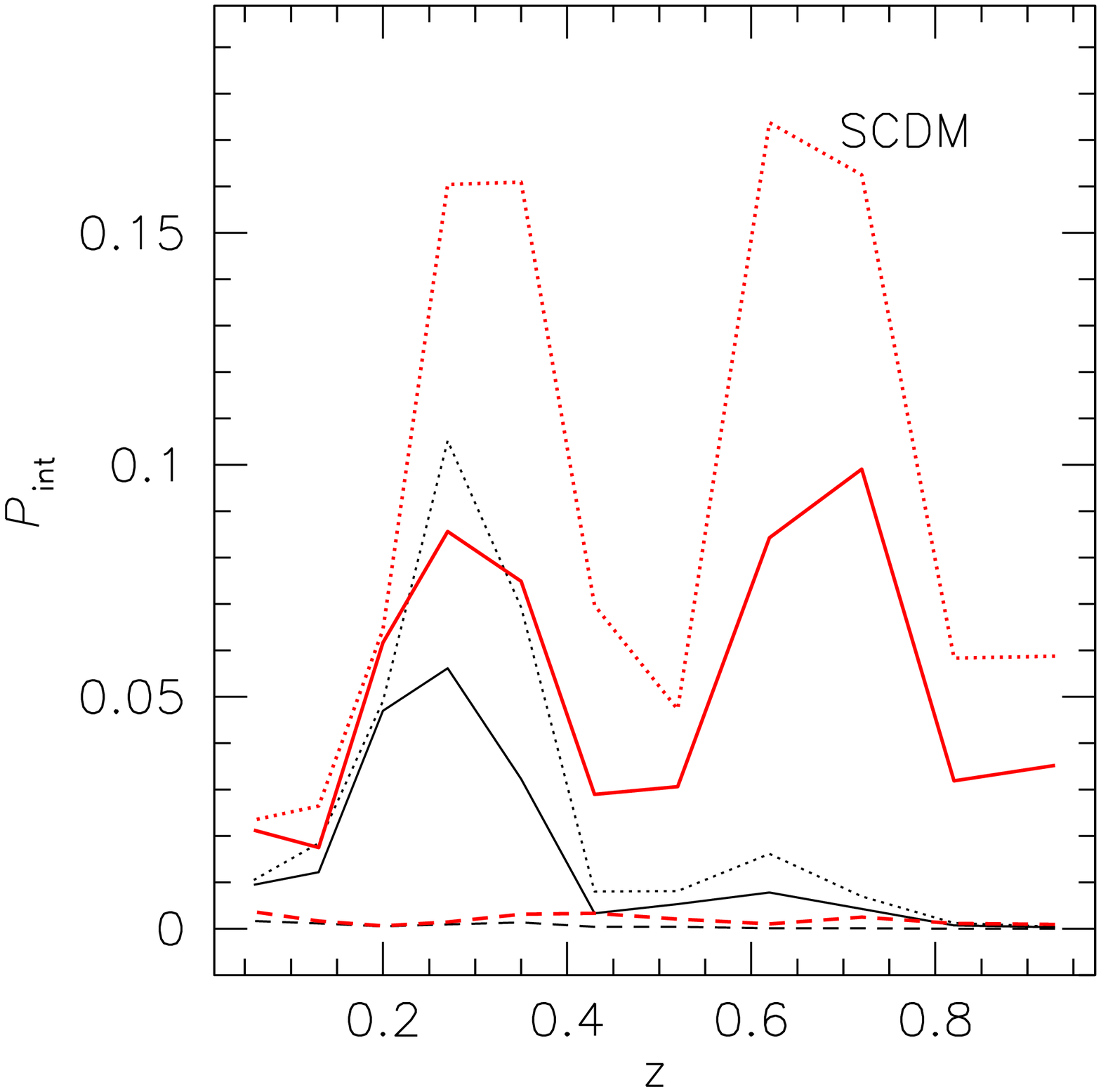,width=.30\textwidth} \hfil
\psfig{file=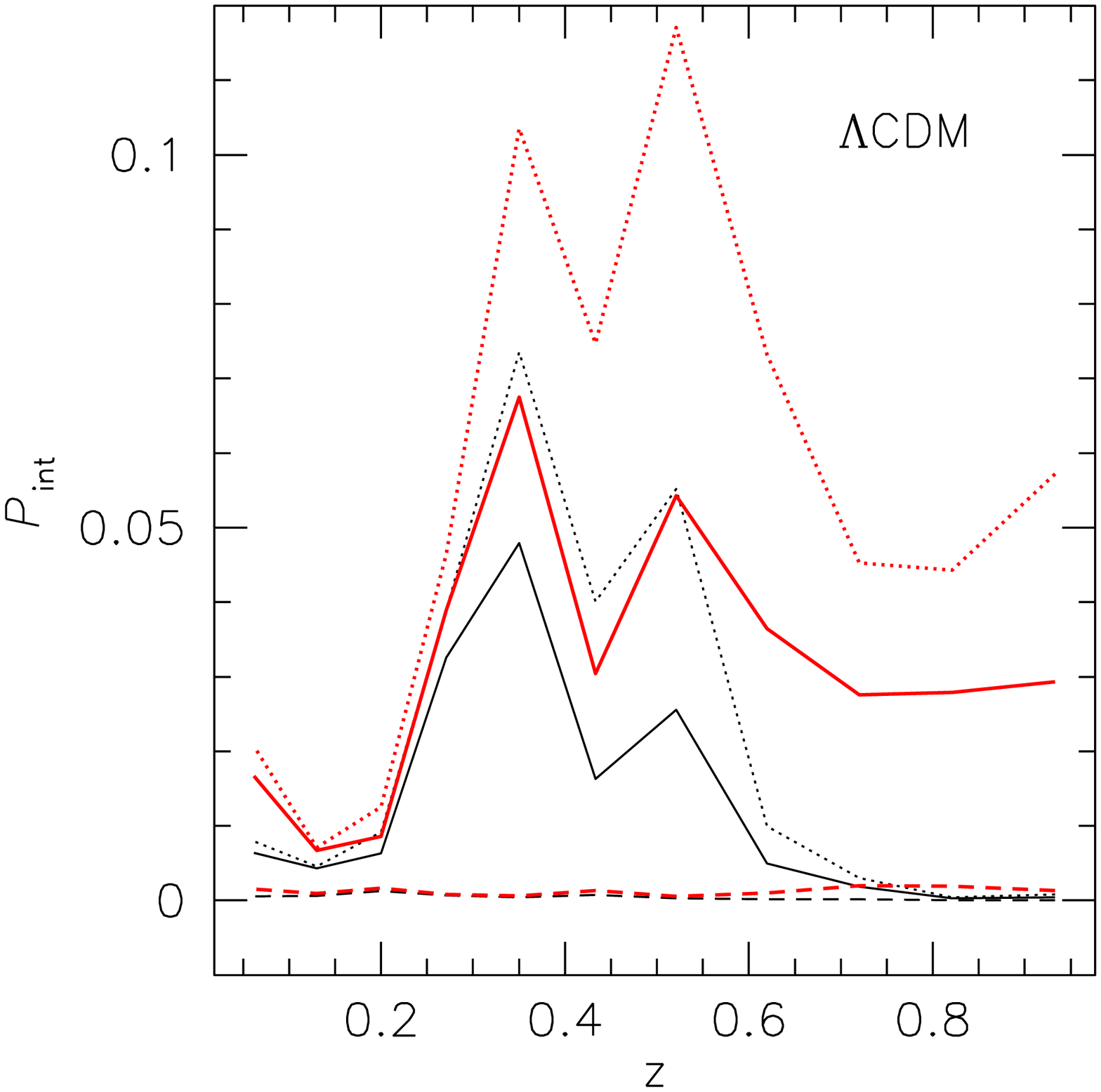,width=.30\textwidth} \hfil
\psfig{file=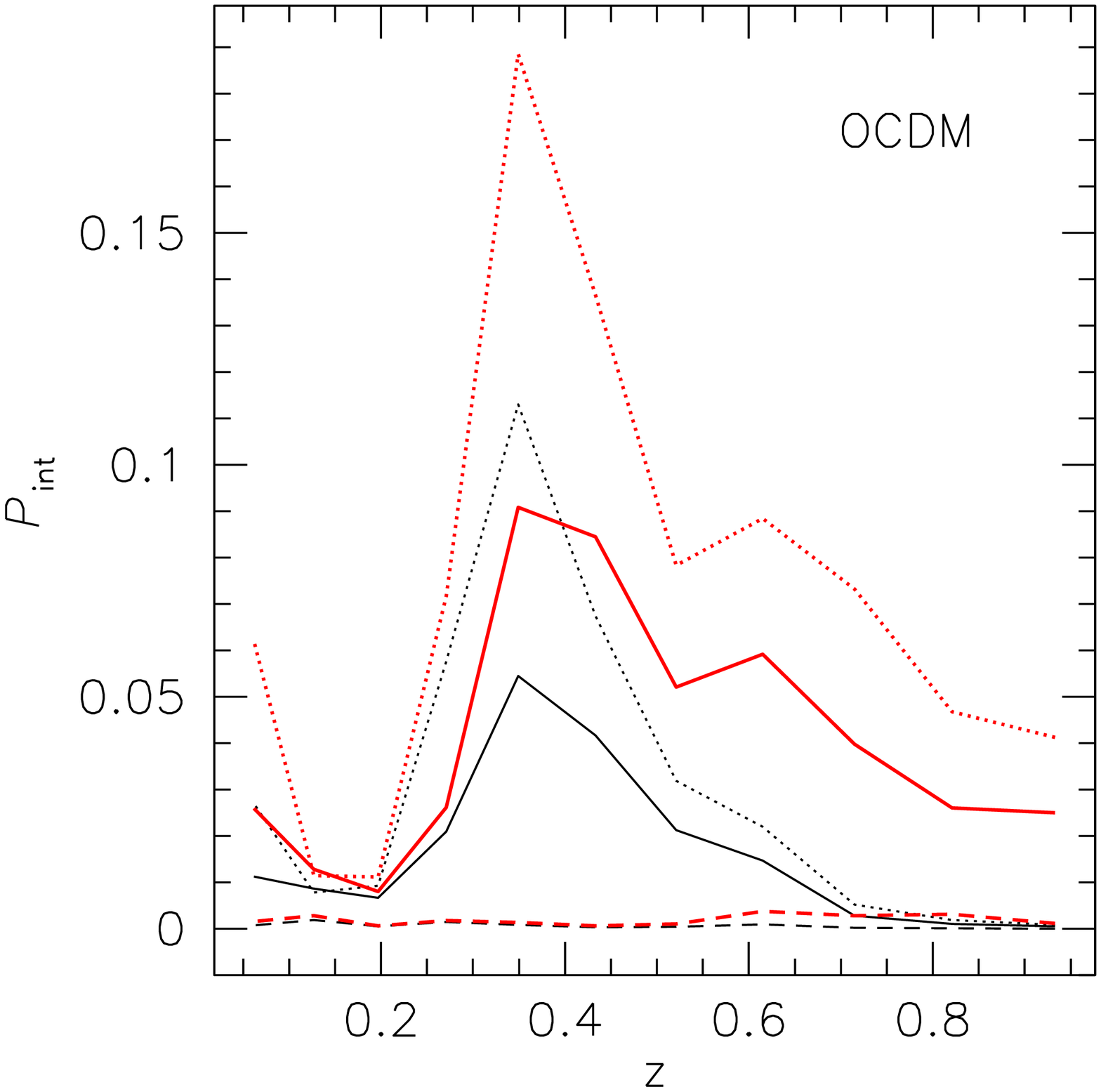,width=.30\textwidth}
}
\caption{Integrated multipole power as a function of redshift for the
  most massive cluster in our numerically simulated sample. Results
  for three different cosmologies are shown; SCDM (left panel),
  $\Lambda$CDM (central panel) and OCDM (right panel). Heavy dotted
  and dashed lines show the results of multipole expansions of the
  cluster surface mass density at cluster-centric distances of
  $r_1=1.25\,h^{-1}{\rm Mpc}$ and $r_0=155\,h^{-1}{\rm kpc}$,
  respectively, and heavy solid lines show the area-weighted averages
  of the integrated power within $r_1$ (see text for more
  details). Light lines show the results of rescaling the heavy curves
  with the cluster virial masses and the effective lensing distance.}
\label{figure:power}
\end{figure*}

Deviations of the projected mass distribution of a numerically
simulated cluster from the predictions of circular or elliptical
models can be quantified by means of a multipole expansion of its
surface density field. For performing this analysis, we first define a
reference frame whose origin coincides with the surface density peak
of the cluster. Then, starting from the particle positions in the
$N$-body simulations, we compute the surface density at discrete radii
$r_n$ and position angles $\phi_k$, with $r_n$ and $\phi_k$ taken from
the intervals $[0,1.5]\,h^{-1}{\rm Mpc}$ and $[0,2\pi]$,
respectively. For any given $r_n$, each discrete sample of data
$\Sigma(r_n,\phi_k)$ can be expanded into a Fourier series in the
position angle,
\begin{equation}
  \Sigma(r_n,\phi_k)=\sum_{l=0}^\infty\,S_l(r_n)
  {\rm e}^{-{\rm i}l\phi_k}\;,
\label{eq:expansion}
\end{equation}
where the coefficients $S_l(r_n)$ are given by
\begin{equation}
  S_l(r_n)=\sum_{k=0}^\infty\,\Sigma(r_n,\phi_k)
  {\rm e}^{{\rm i}l\phi_k}\;,
\end{equation}
and can be computed using fast-Fourier techniques. We define the power
spectrum $P_n(l)$ of the multipole expansion $l$ as
$P_n(l)=|S_l(r_n)|^2$.

Axially symmetric and elliptical models have very simple multipole
expansions. For a circular mass distribution, only the monopole,
$l=0$, contributes to the sum in Eq.~(\ref{eq:expansion}). An
elliptical mass distribution has one more contribution from the
quadrupole, $l=2$. The dipole term, $l=1$, is zero for both these
models. Of course, the surface density fields of numerically simulated
clusters are much more complex than those of our analytic
models. Their multipole expansions contain a dipole and also
multipoles of higher order than two, which correspond to substructures
spanning opening angles of order $\sim\pi/l$ when seen from the
cluster centre.

Therefore, in order to quantify the amount of substructure and the
degree of asymmetry in the mass distributions of our numerically
simulated lenses at any distance $r_n$ from the cluster centre, we can
use the power spectra $P_n(l)$. In particular, we define an integrated
power $P_{\rm int}(r_n)$, which is the sum of the power spectral
densities of the dipole and of all multipoles of higher order than
two, i.e.~we subtract the monopole and quadrupole contributions from
the total integrated power,
\begin{equation}
  P_{\rm int}(r_n)=\sum_{l=0}^\infty\,P_n(l) - P_n(0) - P_n(2)\;.
\end{equation}
This quantity measures the deviation from an elliptical distribution
of the surface mass density at a given distance $r_n$ from the cluster
centre. In order to suppress the dependency on the radial coordinate,
we also compute the area-weighted averaged value of $P_{\rm int}$
inside radius $r$ as
\begin{equation}
  \bar{P}_{\rm int}(r)=\frac{2}{r^2}\,\sum_{r_n\le r}\,
  P_{\rm int}(r_n)\,r_n\Delta r_n\;.
\end{equation}

In Fig.~\ref{figure:power}, we show how the integrated power
$\bar{P}_{\rm int}(r)$ changes as a function of the cluster redshift
for those lenses whose cross sections were plotted in
Fig.~\ref{figure:comparison}. Again, we show the results for all three
cosmological models. For making the integrated power at different
radii comparable, we normalize them to the power of the corresponding
monopole. Moreover, we average the results obtained for the three
projections of the same cluster. The heavy dotted and dashed lines
show the results for radii $r_1=1.25\,h^{-1}{\rm Mpc}$, comparable to
the cluster virial radius, and $r_0=155\,h^{-1}{\rm kpc}$. For
comparison, the heavy solid curves indicate the area-weighted power
inside circles of radius $r_1$ as a function of redshift. The small
values reached by the heavy dashed curves indicate that in the very
central region of the cluster the mass distribution is dominated by
the monopole and the quadrupole terms, i.e.~in the innermost regions
of the clusters, the surface mass density has elliptical
iso-contours. Moreover, these curves are flat, which means that the
surface density contours remain elliptical at all redshifts between
zero and unity. On the other hand, as shown by the heavy dotted and
solid lines, at distances comparable to the virial radius or smaller
the contribution to the power from the dipole and the multipoles of
higher order than two is large and can even exceed $15\%$ of the
monopole contribution.

A quick comparison of these curves to the lensing cross sections in
Fig.~\ref{figure:comparison} shows that the redshifts where the
contributions of the dipole and the higher-order multipoles are
largest correspond quite well to those where the numerical cross
sections deviate most strongly from those of the elliptical models.

This is most obvious between redshifts 0.2 and 0.4, where also the
geometrical lensing efficiency is largest. At lower and higher
redshifts, the lensing cross sections drop for two major reasons;
first, the lenses are too close to the observer or to the sources;
second, the cluster virial mass decreases with increasing
redshift. For better comparing the dependences on redshift of the
integrated power and the lensing cross sections, we rescale
$\bar{P}_{\rm int}$ with the effective lensing distance, $D=D_{\rm l}
D_{\rm ds}/D_{\rm s}$ and with the virial cluster mass.
%
%
The thin lines in Fig.~\ref{figure:power} show $\bar{P}_{\rm int}$
after rescaling. The damping effect almost completely removes the
peaks in the integrated power at redshifts higher than $z\sim0.5-0.6$,
as well as those at redshifts lower than $z\sim0.2$, and make the
curves much more similar to the curves displaying the lensing cross
sections as a function of redshift in Fig.\ref{figure:comparison}.


This correlation between the higher-order multipoles of the mass
distributions and the cross sections proves that substructures and
aymmetries are the dominant reason for the discrepancy between
numerical and analytical lensing cross sections. We verified that the
peaks in the integrated power $\bar{P}_{\rm int}$ occur when clumps of
matter enter within the virial radius of the respective
clusters. Then, the lensing cross section grows for two reasons;
first, because the cluster mass increases, and second, because of the
increased shear produced by the substructures falling towards the
cluster centres.

\section{Summary and discussion}
\label{section:results}

In this paper, we have introduced an analytic lens model for
describing strong lensing by galaxy clusters. It improves upon
previously used models in four ways. First, the commonly adopted
singular isothermal density profile is replaced by the NFW profile,
which is much more adequate for cluster-sized haloes. Second, this
model allows the effect of cosmology on the lens concentration to be
included. Third, we elliptically perturb the model for approximating
the substantial effect of cluster asymmetries on their strong-lensing
cross sections. Fourth, we take into account that sources are
intrinsically elliptical rather than circular. We adapt the
ellipticity of this lens model to fit numerically simulated clusters
and compare the strong-lensing cross sections of the analytic model to
those of fully numerically simulated clusters. Results for singular
isothermal sphere lenses are given for comparison. Our results can be
summarised as follows:

\begin{itemize}

\item The cross section of the axially symmetric NFW lens model for
  arcs with length-to-width ratio larger than $7.5$ and $10$ are
  almost two orders of magnitude smaller than those of simulated
  clusters.

\item The axially symmetric singular isothermal lens model is more
  efficient for strong lensing because its density profile is steeper
  in the core, but its large-arc cross sections are still smaller than
  the numerical ones by typically an order of magnitude, except at low
  redshift where both the numerical models and the NFW model fail in
  producing any lensing effect.

\item The comparatively flat lensing potential of haloes with the NFW
  density profile makes NFW lenses poor image splitters, but efficient
  magnifiers. In addition, the flat potential renders the lensing
  properties very sensitive to changes in the tidal field, because
  small deformations of the potential can lead to large shifts of the
  critical curves.

\item The cross sections of our new elliptical model increase steeply
  and monotonically as a function of ellipticity. An ellipticity of
  $\approx0.3$ typically increases the cross sections by an order of
  magnitude. However, for the model to reproduce the arc cross
  sections of the numerical clusters, unrealistically high
  ellipticities $e\ga0.5$ are required.

\item Comparing the deflection angle maps to those of elliptical NFW
  lens models with variable $e$, we estimated the ellipticity of the
  lensing potential in the central region of the numerically modelled
  haloes and found typical ellipticities of $e\sim0.3$, substantially
  below values required to solve the discrepancy between analytic and
  numerical lens models.

\item The change of the fully numerical cluster cross sections with
  time exhibits pronounced signatures of merger events. As merging
  sublumps approach the cluster centre, their tidal field markedly
  increases the strong-lensing cross section. We have verified this by
  means of a multipole expansion of the cluster surface density field,
  which shows that larger deviations of the lensing cross section of
  the numerical clusters from the prediction of the analytic models
  arise when the contribution to the surface density power spectra
  from the dipole and from higher-order multipoles of high order. Our
  comparison of elliptical analytic lens models with fully numerical
  models shows that an adequate description of such events is
  necessary for an accurate calculation of arc cross sections.

\end{itemize}

We conclude that even our improved analytic model is unable to
reproduce the strong lensing properties of realistic cluster models,
which we assume the fully numerically simulated haloes to be. We have
seen that the axially symmetric NFW lens model underpredicts the
number of arcs with length-to-width ratio exceeding a given threshold
by approximately two orders of magnitude compared to the fully
numerical results.

At present, the NFW density profile can be considered the most
realistic model profile for cluster haloes (see, however, Moore et
al. 1999; Jing \& Suto 2000). Moreover, it allows to take the effect
of varying halo concentration on the strong lensing efficiency into
account, which is a substantial advantage compared to the commonly
used SIS density profile. Finally, the profile permits analytical
calculations of the relevant lensing properties. Therefore, the NFW
density profile appears ideally suited for constructing analytic
models for strong lensing by clusters, in particular if elliptical
distortions are included, as we have done in our extension of the
model.

However, the differences remaining between the fully numerical and
analytic approaches indicate that analytic modelling is still
insufficient for properly and accurately describing strong lensing by
galaxy clusters. We showed that the most important missing factors are
the presence of substructures within the clusters, and the tidal field
of the surrounding matter distribution. In fact, significant
substructure is abundant in and around numerically simulated
haloes. They enhance the shear field around the clusters, increasing
the length of the critical curves and consequently increase the
probability of forming long arcs.

Of course, the elliptically distorted NFW lens model is adequate for
good qualitative calculations of arc probability, as
Fig.~\ref{figure:comparison} shows. However, in order to derive
precise constraints on cosmology or cluster structure and evolution,
more realistic cluster lens models are required, for which numerical
simulations seem to be the only reliable choice.

\section*{Acknowledgements}  

This work has been partially supported by Italian MIUR (Grant 2001,
prot. 2001028932, ``Clusters and groups of galaxies: the interplay of
dark and baryonic matter''), CNR and ASI.  MM
thanks the EARA for financial support and the Max-Planck-Institut
f\"ur Astrophysik for the hospitality during the visits when part of
this work was done.  We are grateful to Bepi Tormen for clarifying
discussions and to the anonymous referee for useful comments.

\end{document}